\documentclass[aps,prd,preprint,showpacs,preprintnumbers,nofootinbib]{revtex4}
\usepackage{graphicx,epsfig}
\usepackage{latexsym,amsmath,amssymb,amstext}
\usepackage{float}
\usepackage[paperwidth=210mm,paperheight=297mm,centering,hmargin=2cm,vmargin=2.5cm]{geometry}

\newcommand{\be}{\begin{equation}}
\newcommand{\ee}{\end{equation}}
\newcommand{\bea}{\begin{eqnarray}}
\newcommand{\eea}{\end{eqnarray}}

\newcommand{\topo}{{\rm \scriptscriptstyle top}}

\newcommand\bef{\begin{figure}}
\newcommand\eef[1]{\label{fg:#1}\end{figure}}
\newcommand\beq{\begin{equation}}
\newcommand\eeq[1]{\label{#1}\end{equation}}
\newcommand\beqa{\begin{eqnarray}}
\newcommand\eeqa[1]{\label{#1}\end{eqnarray}}
\newcommand\bet{\begin{table}}
\newcommand\eet[1]{\label{tb:#1}\end{table}}

\newcommand\fgn[1]{Figure \ref{fg:#1}}
\newcommand\eqn[1]{Eq.\ (\ref{#1})}

\newcommand\ie{{\sl i.e.\/}}

\begin{document}

\title{Flavor and topological current correlators in parity-invariant three-dimensional QED}

\author{Nikhil\ \surname{Karthik}}
\email{nkarthik@fiu.edu}
\affiliation{Department of Physics, Florida International University, Miami, FL 33199.}

\author{Rajamani\ \surname{Narayanan}}
\email{rajamani.narayanan@fiu.edu}
\affiliation{Department of Physics, Florida International University, Miami, FL 33199.}

\begin{abstract}
We use lattice regularization to study the flow of the flavor-triplet
fermion current central charge $C_J^f$ from its free field value
in the ultraviolet limit to its conformal value in the infrared
limit of the parity-invariant three-dimensional QED with two flavors
of two-component fermions. The dependence of $C_J^f$ on the scale
is weak with a tendency to be below the free field value at
intermediate distances.  Our numerical data suggests that the
flavor-triplet fermion current and the topological current correlators
become degenerate within numerical errors in the infra-red limit,
thereby supporting an enhanced O$(4)$ symmetry predicted by strong
self-duality.  Further, we demonstrate that fermion dynamics is
necessary for the scale-invariant behavior of parity-invariant
three-dimensional QED by showing that the pure gauge theory with
non-compact gauge action has non-zero bilinear condensate.

\end{abstract}

\date{\today}
\maketitle

\section{Introduction}

There is significant numerical evidence that parity invariant three
dimensional QED is scale-invariant for all even values of $N$, the
number of flavors of massless two-component
fermions~\cite{Karthik:2015sgq,Karthik:2016ppr}. In particular, it
has been shown that the theory with $N=2$ is consistent with a
vanishing bilinear condensate using two different lattice regularization
schemes. It is now important to characterize the infra-red fixed
point for $N=2$.

Denoting the two flavors of two-component fermions by
$(\bar\chi_i,\chi_i)$, $i=1,2$, we define one of the flavor-triplet
scalar and vector bilinear operators as
\be
\Sigma({\bf x}) = \bar\chi_1({\bf x}) \chi_1({\bf x}) - \bar\chi_2({\bf x})\chi_2({\bf x});\qquad
V_k({\bf x}) = \bar\chi_1({\bf x})\sigma_k  \chi_1({\bf x}) - \bar\chi_2({\bf x})\sigma_k \chi_2({\bf x}),
\ee
where $\mathbf{x}=(x,y,t)$. In~\cite{Karthik:2016ppr}, we showed
that both the correlators show massless behavior, and we provided
some results concerning the scaling dimensions of these two operators.
The scalar correlator gradually changes from the free field behavior
of $|\mathbf{x}|^{-4}$ at short distances to
$|\mathbf{x}|^{-2\Delta_\Sigma}$ at large distances and the scaling
dimension was found to be $\Delta_\Sigma = 1.0 \pm 0.2$.  This
result is consistent with the one obtained
in~\cite{Rantner:2000wer,Rantner:2002zz} using an expansion in large
number of flavors~\footnote{Care should be used in taking this
agreement at face value since an agreement is found by setting the
number of flavors to two in their computation which need not be
large.} and with our own estimate of the mass anomalous dimension
from the finite size scaling of the low-lying eigenvalues of the
Dirac operator.  The power-law decay of the flavor-triplet vector
correlator remains $|\mathbf{x}|^{-4}$ at all distances since it
is a conserved current. Since, we project correlators to zero spatial
momentum to study them as a function of the Euclidean time separation
$t$, the vector correlator decays as $t^{-2}$ and the coefficient,
$C^f_J(t)$, of this power-law decay~\footnote{Since we are interested
in ratios of $C_J^f$, any difference by a factor in our definition
of $C_J^f$ from elsewhere in the literature is inconsequential.}
is what we refer to as the amplitude, and it becomes the flavor
current central charge at the conformal point in the infra-red limit
$t\to\infty$.

In this paper,  we extend our results further in the following three ways:
\begin{enumerate}
\item 
Assuming conformal symmetry that is valid for large number of flavors
and using a diagrammatic approach~\cite{Huh:2014eea,Giombi:2016fct},
the amplitude of the correlator of the vector bilinear is found to
be
\be
\frac{C^f_J(t\to\infty)}{C^f_J(t\to0)} = 1 + \frac{0.1429}{N} + \mathcal{O}\left(\frac{1}{N^2}\right). \label{giombiv}
\ee
Thus, in the infra-red limit the value of $C_J^f$ is larger the the
free field value in the ultra-violet by a factor 1.07 for $N=2$.
Numerical conformal bootstrap~\cite{Chester:2016wrc} has been used
to obtain the allowed region for this amplitude directly for two flavors.
In this work, we study the behavior of $C_J^f(t)$.  For the values
of $t$ where a reliable numerical estimate is possible our value
lies close to its ultraviolet value. However, we find that $C_J(t)$
has a tendency to flow from its ultraviolet value at small $t$
to a value below it at intermediate values of $t$. If it has to
agree with the result from the diagrammatic approach in \eqn{giombiv},
the flow has to be non-monotonic, and our result does not strongly
support it.
\item A self-duality has been proposed to be valid at the
infra-red fixed point of $N=2$ two component QED~\cite{Xu:2015lxa,Karch:2016sxi,Hsin:2016blu,Wang:2017txt}.
Since the topological current on one side of the duality maps onto
the flavor-triplet vector current on other side of the duality,
their correlators have to be degenerate at large separations. This
also implies that the amplitude of the correlator of the vector
bilinear $C_J^f$ and the amplitude of the topological current
correlator $C_J^t$ have to be the same. This SU$(2)\times$ SU$(2)$ symmetry becomes an
emergent O$(4)$ symmetry~\cite{Wang:2017txt}. We provide evidence in favor of this
argument. This would imply that the infra-red fixed point in QED$_3$
coupled to small number of fermion flavors is qualitatively different
from the one expected in large $N$.

\item Unlike the theory with two flavors of two component fermions,
quenched QED (limit where the number of flavors is taken to zero)
has a non-zero bilinear condensate.  This can be considered as a
follow-up of a calculation~\cite{Hands:1989mv} done three decades
ago when computational power was not sufficient to extract the
continuum value of the condensate.  Thus, the fermions used as a
probe in pure gauge theory develops a scale, and fermion dynamics
is necessary for a scale-invariant behavior.

\end{enumerate}

\section{Flow of $C_J^f$ from the ultraviolet to the infrared}

\bef
\centering
\includegraphics[scale=1.1]{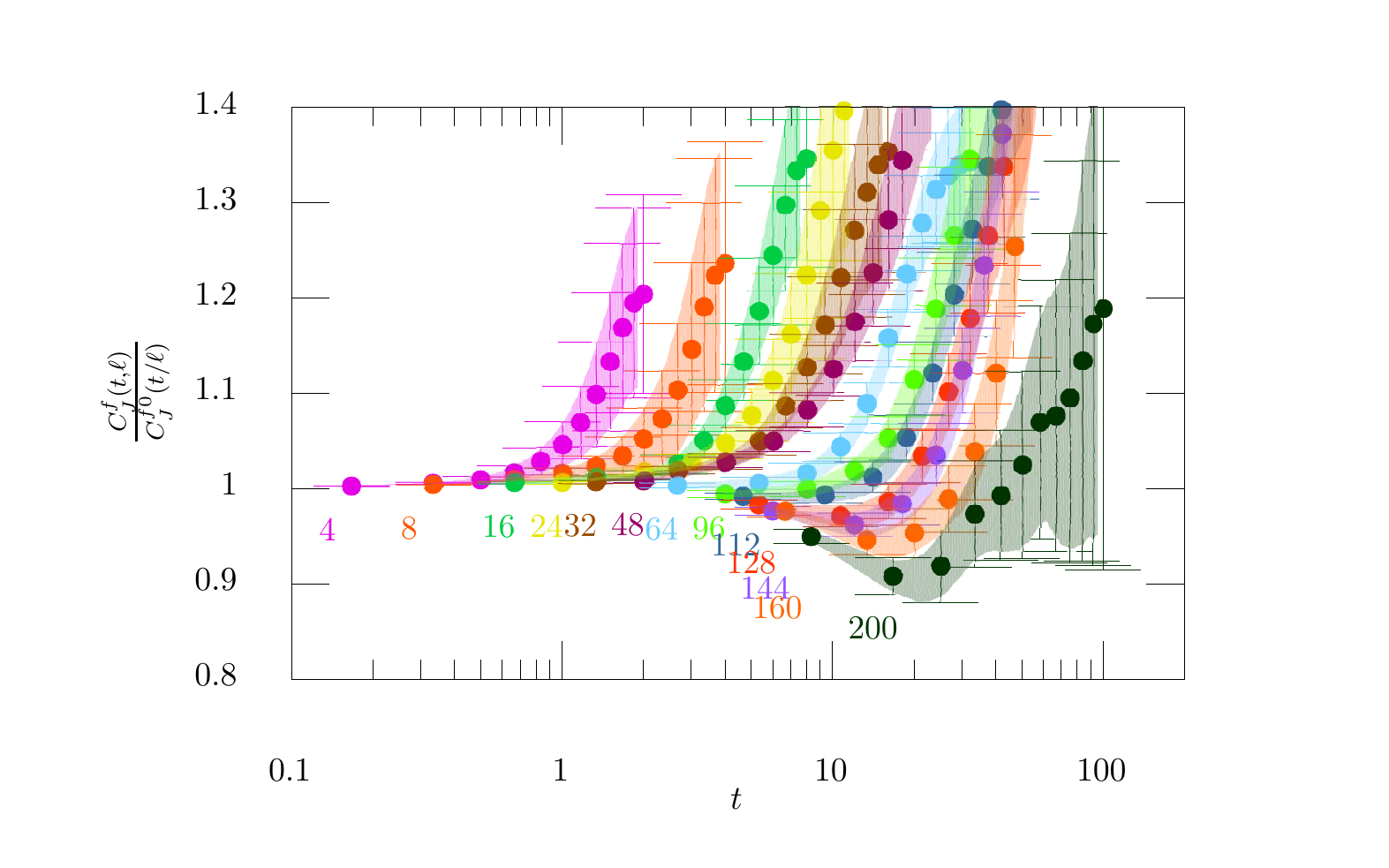}
\caption{The data for $\frac{C_J^f(t,\ell)}{C_J^{f0}(t/\ell)}$ from different $\ell$ is put together as a function of $t$. All the correlators 
in the figure were determined on finite lattice $L=24$. The 1-$\sigma$ error bands interpolating the data points are also shown.}
\eef{cjdata}

We simulated $N=2$ QED$_3$ at different finite physical volumes
$\ell^3$ regulated on lattice with $L$ points in each direction.
The details of the simulation are given in~\cite{Karthik:2016ppr}.
We analyzed the data at $L=12,14,16,20,24$ and
$\ell=4,8,16,24,32,48,64,96,112,128,144,160,200$.  In order to
improve the signal, we project the correlators to zero momentum in
spatial directions.  The zero spatial momentum projected flavor-triplet
vector correlator determined at finite physical volume is
\be
G_V(t,\ell) = \int dx \ dy\  \left\langle \sum_{k=1}^2 V_k(0,0,0) V_k(x,y,t)\right\rangle \equiv \frac{C^f_J(t,\ell)}{t^2}.
\label{vecdef}
\ee
The corresponding expression on the lattice in terms of the overlap
fermion propagators is given in~\cite{Karthik:2016ppr}.  In order
to study the flow of $C_J^f$ from UV to IR, we study the ratio
$C_J^f(t,\ell)$ in the interacting theory to the free field value
$C_J^{f0}\left(\frac{t}{\ell}\right)$ obtained on the same $L^3$
lattices \ie,
\be
\frac{C_J^f(t,\ell)}{C_J^{f0}(\frac{t}{\ell})} = \frac{G_V(t,\ell)}{G_V^{\rm\scriptscriptstyle free}(\frac{t}{\ell})},
\ee
where $G_V^{\rm\scriptscriptstyle free}$ is the correlator obtained by putting all lattice gauge fields to zero.

In order to obtain the ratio $\frac{C_J^f}{C_J^{f0}}$ in the continuum
limit as well as in the infinite volume limit, one has to take the
$L\to\infty$ limit of the ratio at different $t$ at fixed $\ell$,
and then take the $\ell\to\infty$ limit at fixed $t$. Before we
incorporate this procedure, we put together the data for the ratio
from different $\ell$ at $L=24$ as a function of $t$ in \fgn{cjdata}.
At finite $L$ and $\ell$, we only obtain values for
$G_V\left(t,\ell\right)$ at certain discrete values $t=T\frac{\ell}{L}$
where $T=1,\ldots,\frac{L}{2}$ --- these are the solid circles in
\fgn{cjdata}, with each color corresponding to data from different
$\ell$. We can qualitative see the following. At small $t$, the
value of $\frac{C_J^f}{C_J^{f0}}$ is almost unity as expected.
However, at any larger fixed value of $t$, the value of the ratio
decreases with $\ell$ and goes below unity for certain intermediate
$t$.

\bef
\centering
\includegraphics[scale=0.82]{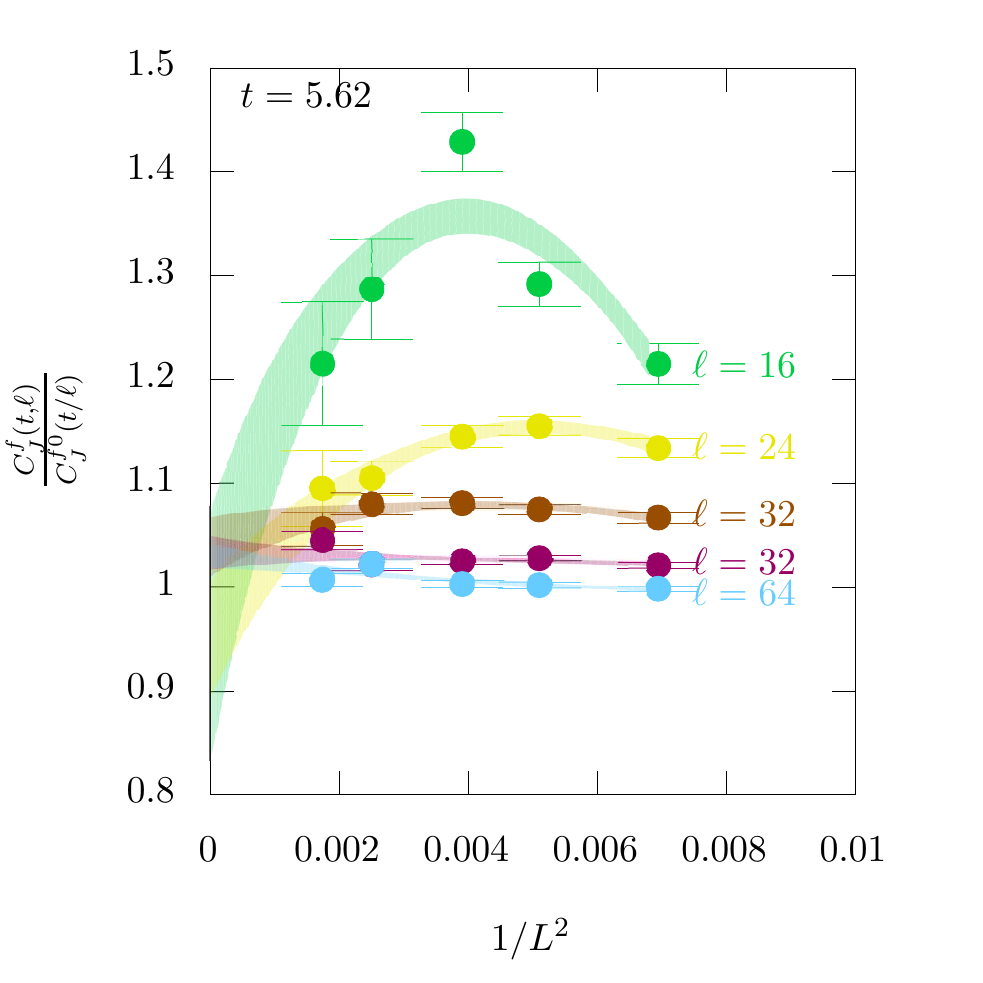}
\includegraphics[scale=0.82]{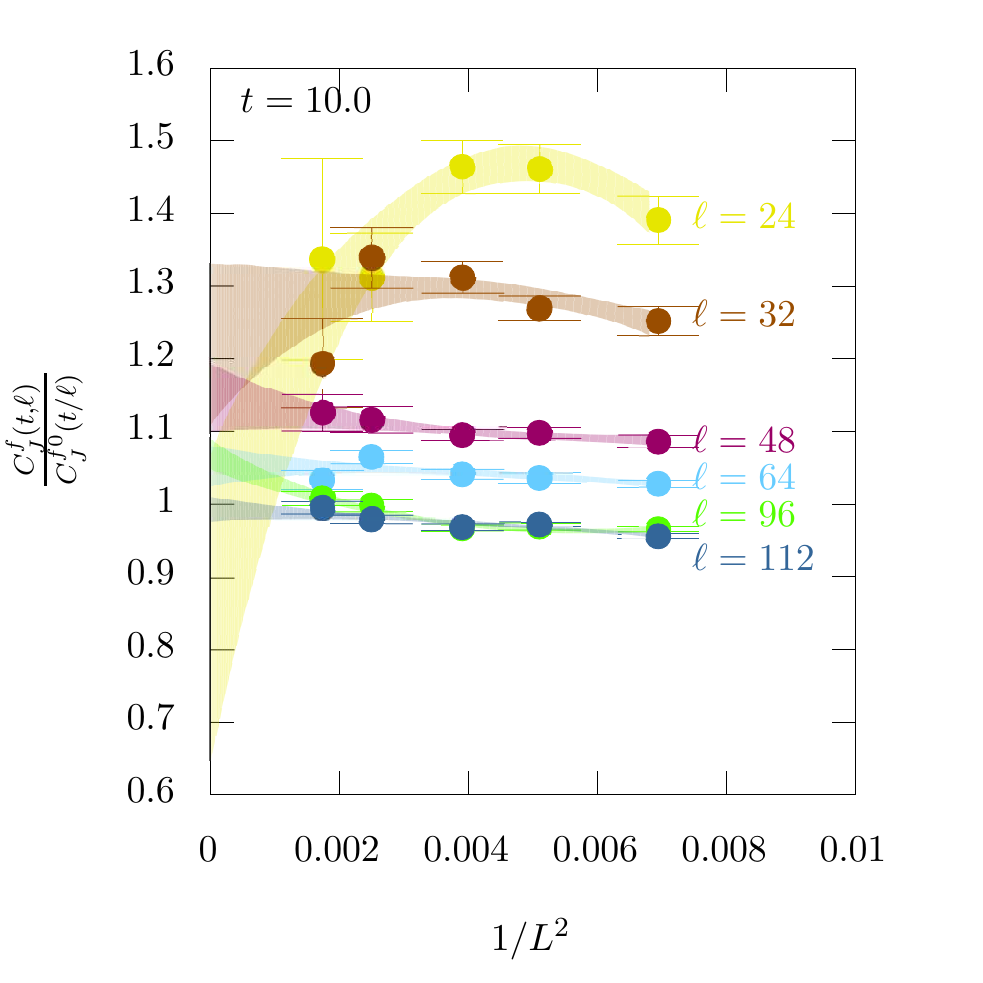}

\includegraphics[scale=0.82]{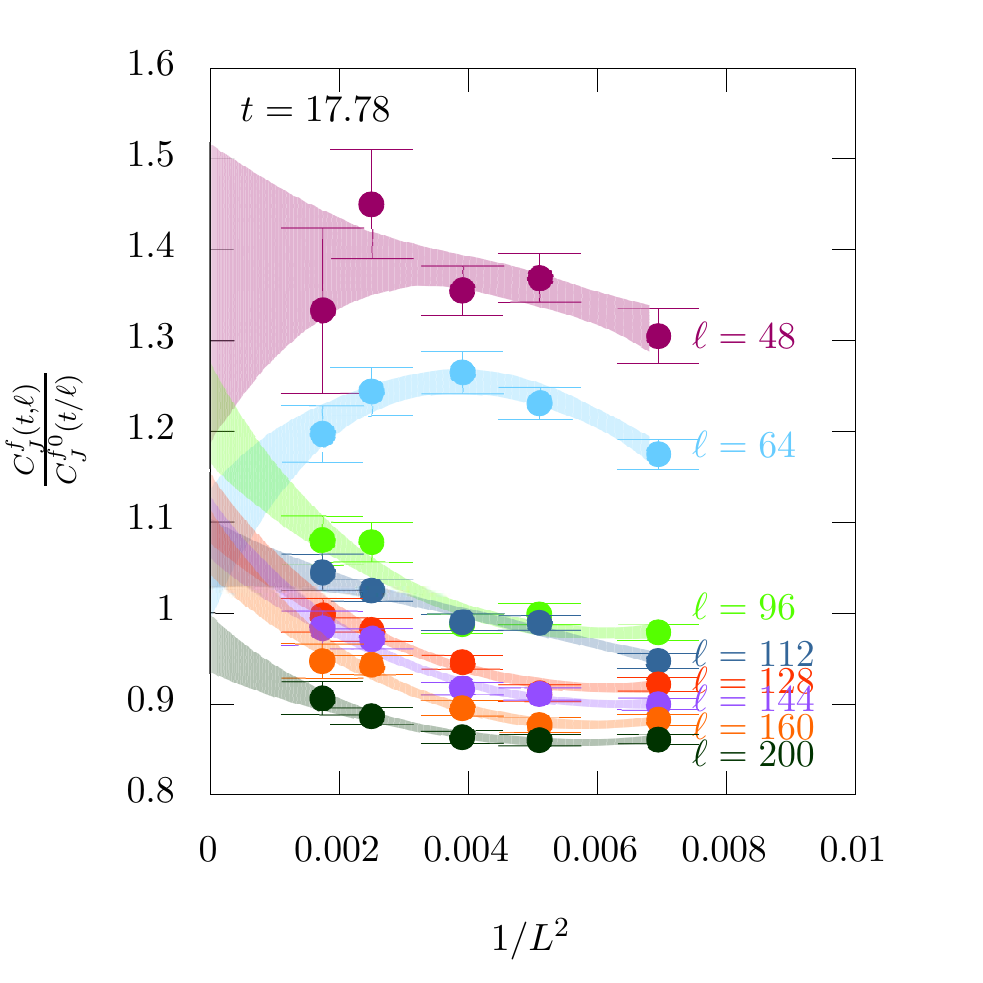}
\includegraphics[scale=0.82]{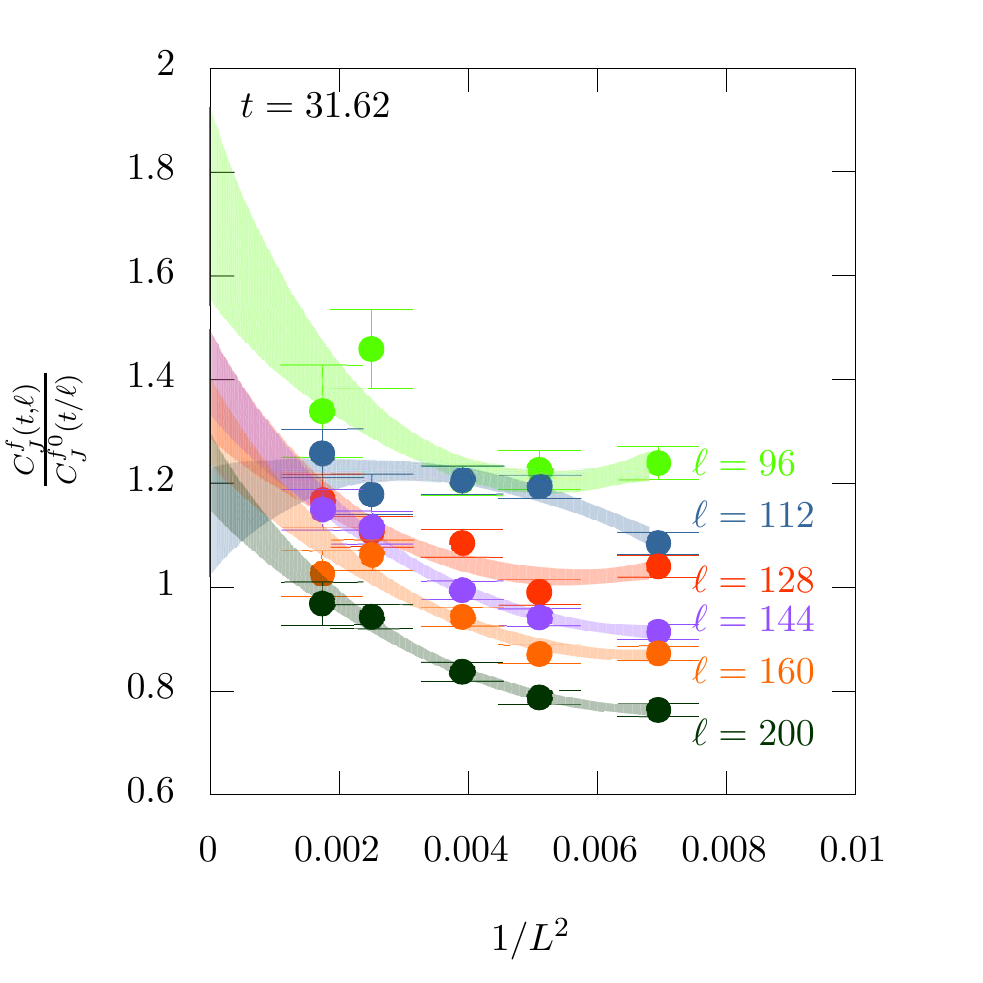}
\caption{Continuum limits, $\lim_{L\to\infty}\frac{C_J^f(t,\ell)}{C_J^{f0}(t/\ell)}$ 
at various $\ell$ (differentiated by the colors in each panel) 
at the same values of $t$ using $L=12,14,16,20$ and 24. The four panels correspond to different 
$t$. The continuum extrapolation is using a $A+B/L^2+C/L^3$ fit to the data. 
The 1-$\sigma$ error bands for the extrapolation are shown along with the data.}
\eef{cjcontinuum}
\bef
\centering
\includegraphics[scale=0.82]{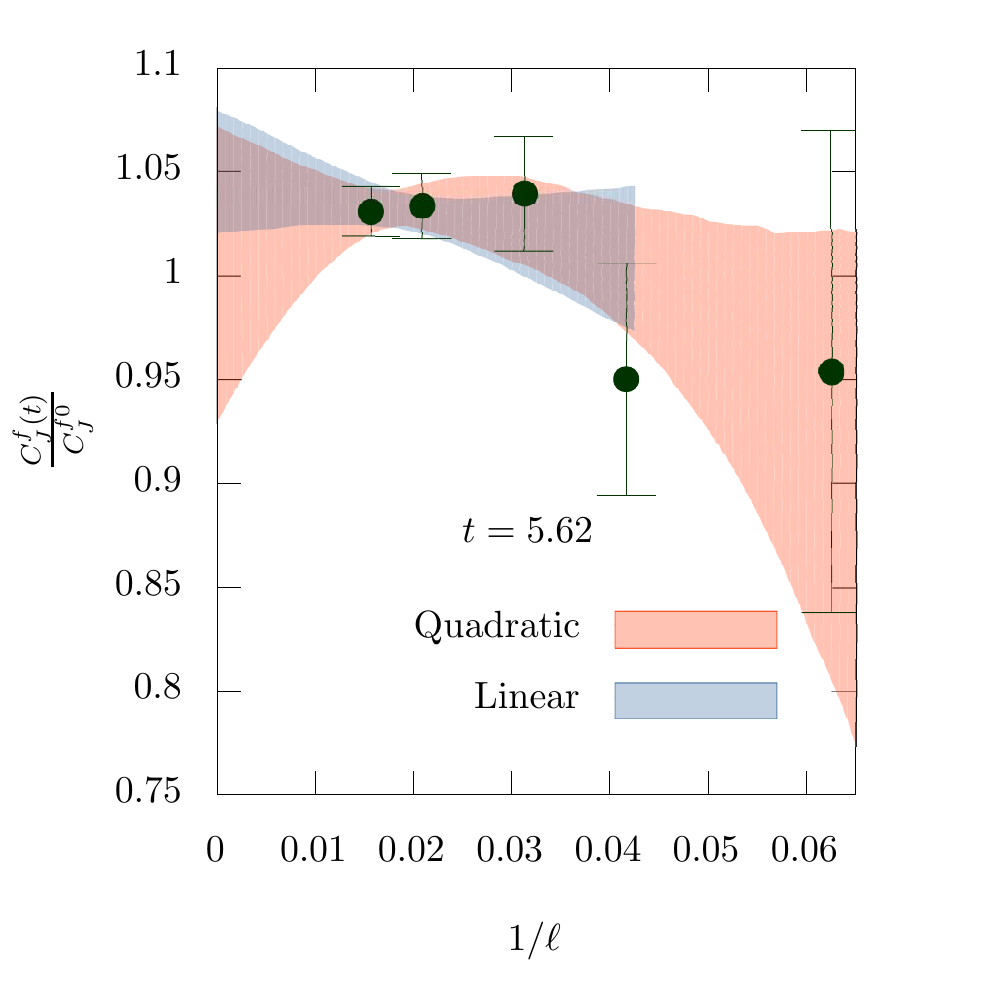}
\includegraphics[scale=0.82]{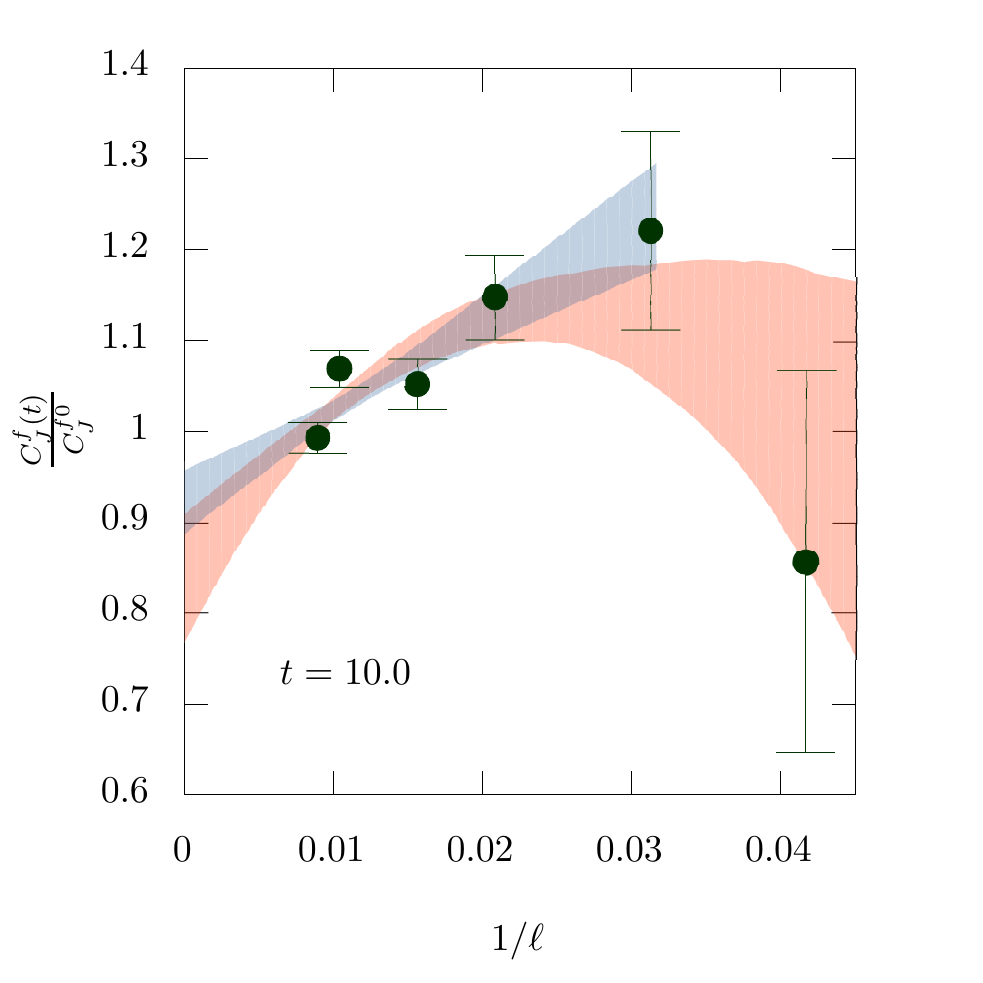}

\includegraphics[scale=0.82]{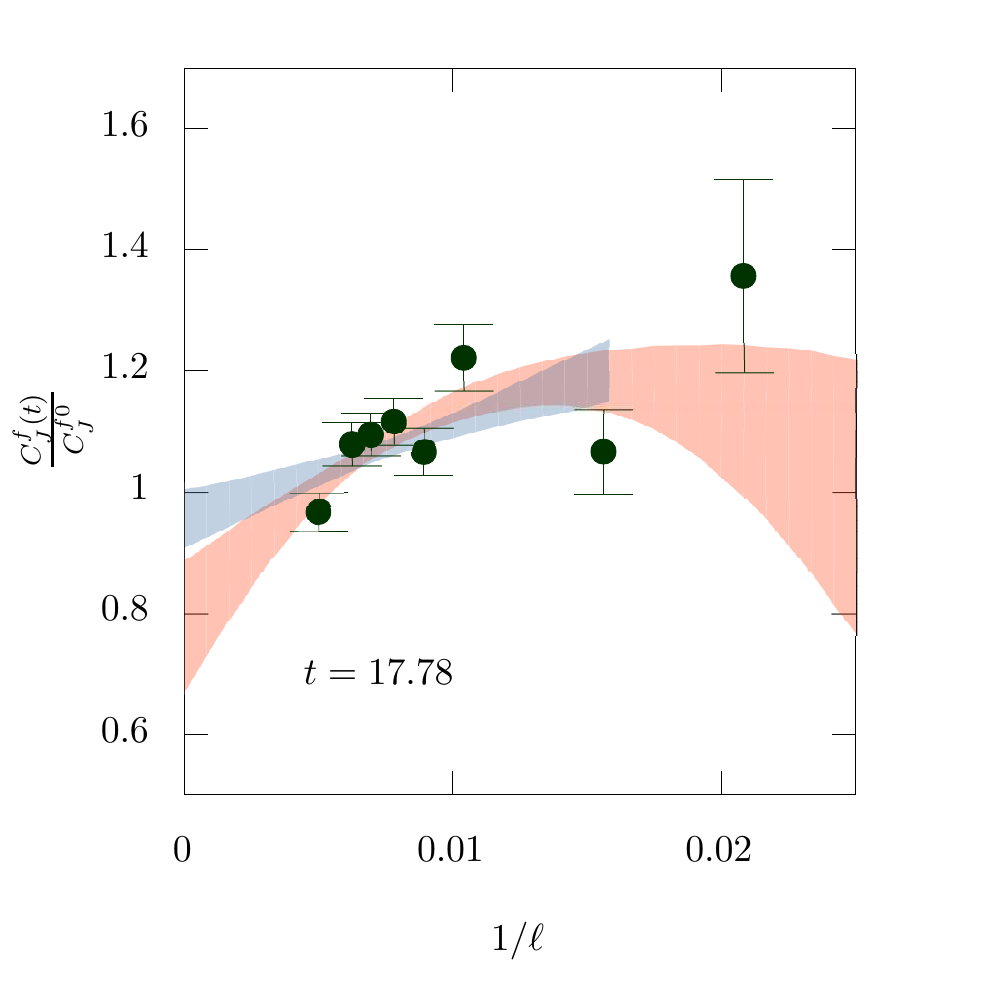}
\includegraphics[scale=0.82]{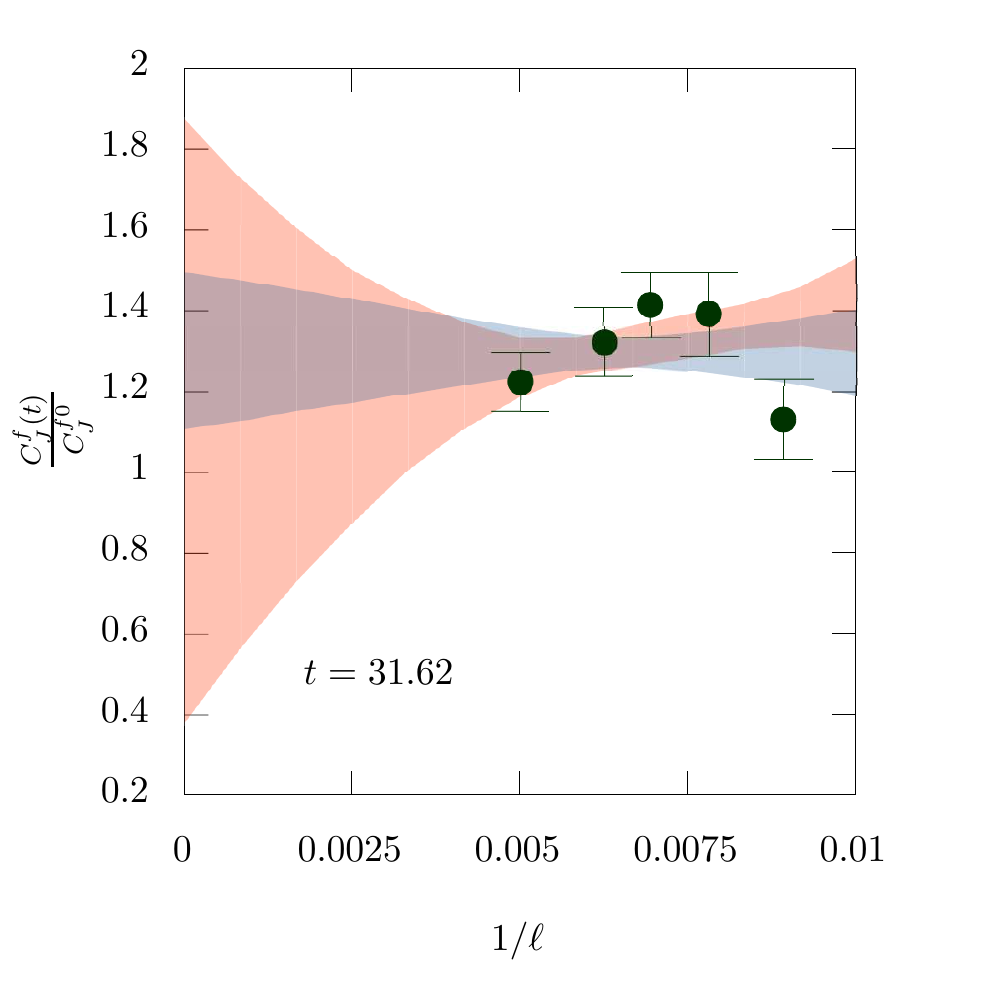}
\caption{Infinite volume limits,
$\lim_{\ell\to\infty}\frac{C_J^f(t,\ell)}{C_J^{f0}(t/\ell)}$, at
different fixed values $t$ are shown in the four panels.  The black
circles are the values obtained in the continuum limit (refer
\fgn{cjcontinuum}) at different $\ell$ at a value of $t$. The
1-$\sigma$ error band for~$\frac{C_J^f}{C_J^{f0}}\left(t\right)+\frac{k_1}{\ell}$ extrapolation
is shown in red. The 1-$\sigma$ error band for~$\frac{C_J^f}{C_J^{f0}}\left(t\right)+\frac{k_1}{\ell}+\frac{k_2}{\ell^2}$
extrapolation is shown in blue.}
\eef{cjinfvol}

Now we proceed to take care of the finite lattice spacing and the
finite volume effects in the data. First, we interpolate our data
between the discrete values of $t=T\frac{\ell}{L}$ using cubic
spline. This is justified since the data for the ratio is smooth
and regular as seen in \fgn{cjdata}. The error bars on the interpolation
is obtained by bootstrap.  The 1-$\sigma$ error band for the interpolation
is shown along with the data in \fgn{cjdata}.  This gives us results
in the range $t\in [\frac{\ell}{L},\frac{\ell}{2}]$.  In \fgn{cjcontinuum},  we
address the lattice spacing effects. Each panel
corresponds to a fixed value of $t$.  Given that we wish to use
data at all five values of $L$ to obtain the continuum limit, we
can only use $\ell$ ranging from $2t$ to  $12t$ at a given $t$.
These are the different colored symbols in each panel in
\fgn{cjcontinuum}. Since we have used fermions with exact flavor
symmetry on the lattice, the leading lattice correction is
$\mathcal{O}\left(\frac{1}{L^2}\right)$, and we include $\frac{1}{L^2}$ and
$\frac{1}{L^3}$ corrections to extrapolate to the continuum limit $L\to\infty$.
These extrapolations are shown by the error bands in \fgn{cjcontinuum}.

Using the continuum limits so obtained for
$\frac{C_J^f\left(t,\ell\right)}{C_J^{f0}\left(\frac{t}{\ell}\right)}$, we
show its $\ell$ dependence at various $t$ in the four panels of
\fgn{cjinfvol}. We were able to capture the $\ell$ dependence by a
linear,~$\frac{C_J^f}{C_J^{f0}}\left(t,\ell\right)=
\frac{C_J^f}{C_J^{f0}}\left(t\right)+\frac{k_1}{\ell}$, dependence
in the range of $t$ we explored. This is shown as the blue 1-$\sigma$ error band in
the different panels. However, to address systematic effects of the
fit, we also use a quadratic
fit,~$\frac{C_J^f}{C_J^{f0}}\left(t,\ell\right)=
\frac{C_J^f}{C_J^{f0}}\left(t\right)+\frac{k_1}{\ell}+\frac{k_2}{\ell^2}$,
to extrapolate to $\ell\to\infty$.  This is shown as the red 1-$\sigma$ error bands in
the panels. At smaller $t$, the errors are smaller and hence the
errors on the extrapolations are controlled. In fact for $t<6$,
only a weak dependence on $\ell$ is seen and one can drop any
$1/\ell$ dependence, and the values are consistent with 1.  But
there exists a range of $t$ ($t=10.0$ falls in this range) where
this quantity has a value less than unity. This does not violate
the requirement of monotonic decrease of the propagator with $t$
since it only implies the relation
\be
\frac{ d \ln C_J^f(t,\infty)}{d t} < \frac{2}{t}.\label{derbound}
\ee
But it suggests that $C_J^f(t,\ell\to\infty)$ cannot be a monotonic
function of $t$ if it has to be consistent with \eqn{giombiv}.
However, as $t$ is increased the errors increase, and hence we lose
our ability to determine the infinite volume limit for $t>30$.

\bef
\centering
\includegraphics[scale=0.9]{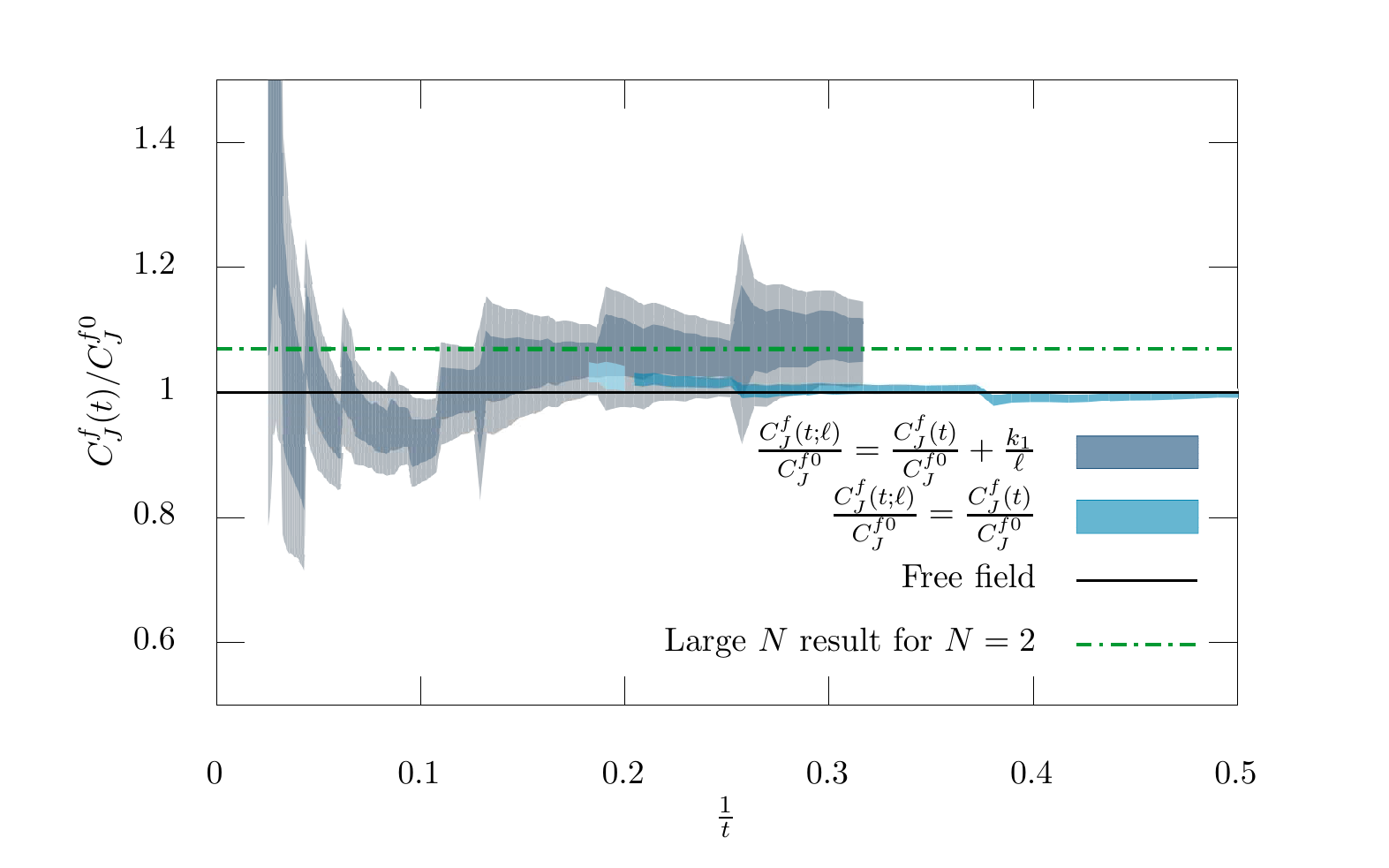}
\includegraphics[scale=0.9]{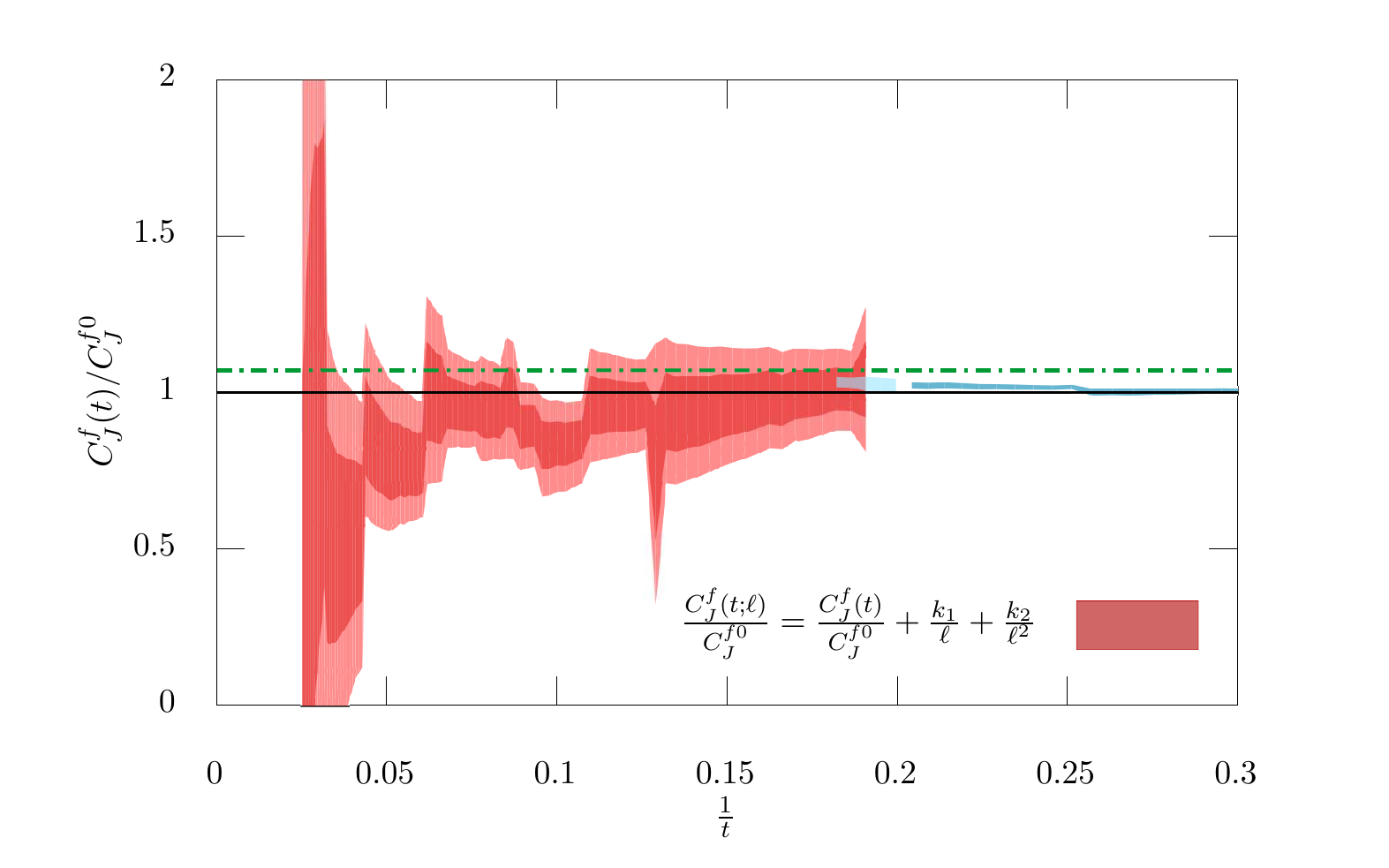}
\caption{Flow of $\frac{C_J^f(t)}{C_J^{f0}}$ from the UV to the IR fixed point. The top and the 
bottom panels differ by the fits used for the $\ell\to\infty$ extrapolation; the top panel includes 
only the $1/\ell$ effect (as shown by the blue bands in \fgn{cjinfvol}) while the bottom panel includes 
$1/\ell$ as well as $1/\ell^2$ effects (as shown by red bands in \fgn{cjinfvol}). The darker
band is the 68\% confidence interval and the lighter band is the
95\% confidence interval. In both the top and the bottom panels, the blue thin band that remains very close to 1 for $1/t>0.2$ is obtained 
assuming no finite $\ell$ effects while taking the $\ell\to\infty$ limit at those $t$. The black line is the free field value in the ultraviolet $1/t\to\infty$ limit while the 
green dotted line is expectation in the infrared $1/t\to 0$ limit from a large $N$ computation.}
\eef{cjflow}

The flow of $C_J^f(t)$ in infinite physical volume from its ultraviolet
value normalized to unity toward its infrared value is shown in
\fgn{cjflow}. The top panel shows the result obtained using a linear
extrapolation in $1/\ell$ to the infinite volume limit at fixed values of $t$.
The darker band shows the $68\%$ confidence interval, while the
lighter band encloses $95\%$ confidence interval. We see that the
flow either remains at the ultraviolet value or it increases slightly
first from its value in the ultraviolet limit.  In the same panel,
the infinite volume limits at $t<6$ obtained assuming no $\ell$
dependence in the data is shown as the light blue band. It is even more
evident that $C_J^f(t)$ approaches the free field value in the
ultraviolet limit.  There is an intermediate region in $t$ (around
$t=10$) where there is evidence that it is below its value in the
ultraviolet limit.  The bottom panel of \fgn{cjflow} compares the
estimate of the flow when both $1/\ell$ and $1/\ell^2$ terms are used to estimate
the value at infinite volume. We see that relevant qualitative
aspects of the flow are not affected by the choice of the fit. In
particular, the inclusion of higher order corrections in $\frac{1}{\ell}$
suggests that the flow remains below free field value even beyond
the intermediate region in $t$.  If this trend continues at even larger
$t$ closer to the infra-red limit, it would be inconsistent with
\eqn{giombiv}, but that is a result valid for large number of
flavors. An analytic calculation at finite $N$ for the flow of $C_J^f$ 
near the infra-red fixed point (\ie, large but finite $t$) 
would enable an extrapolation of our result, reliable at finite $t$, to $t\to\infty$.

\section{Enhanced O$(4)$ symmetry}

\bef
\centering
\includegraphics[scale=0.6]{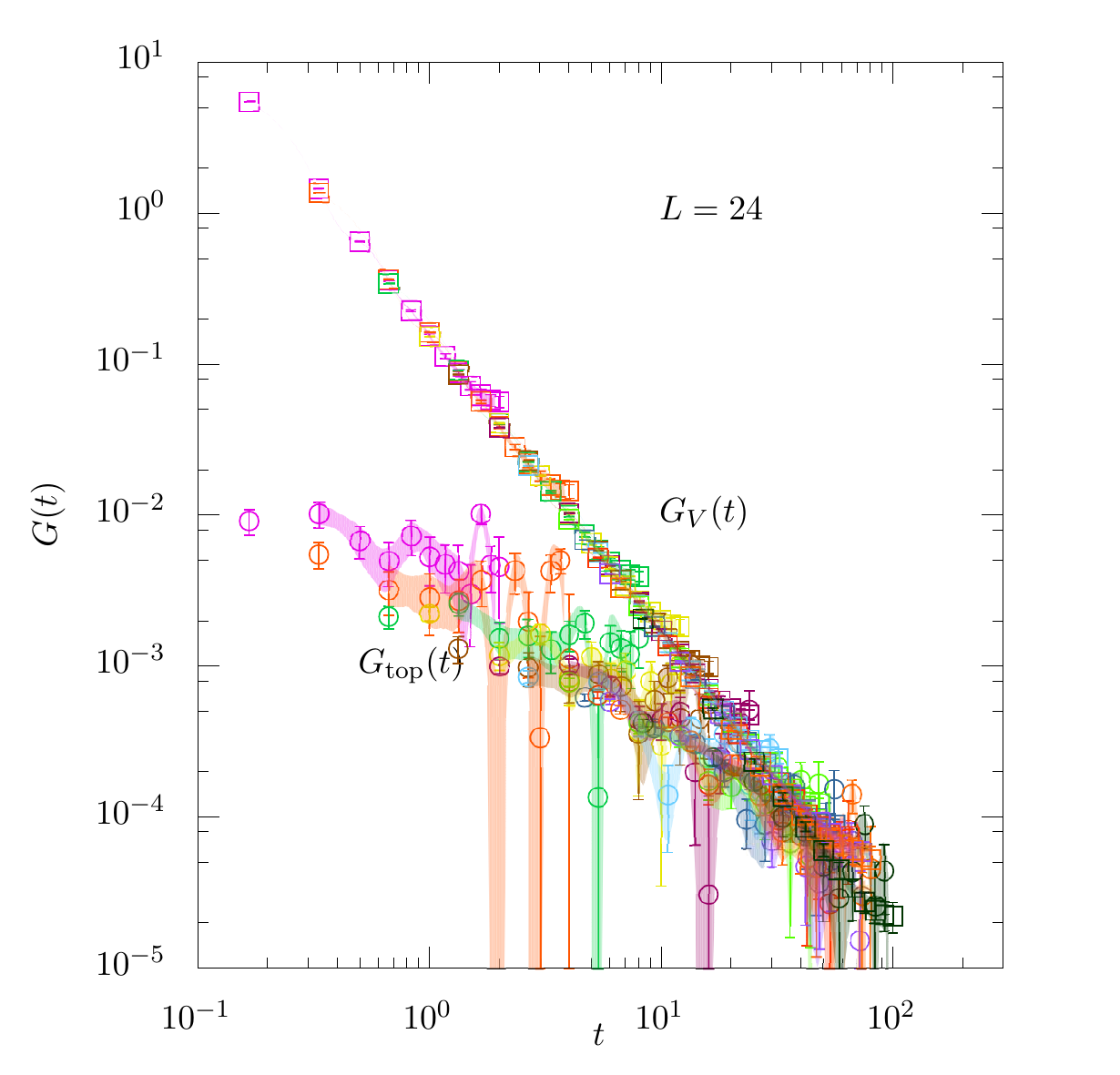}
\includegraphics[scale=0.6]{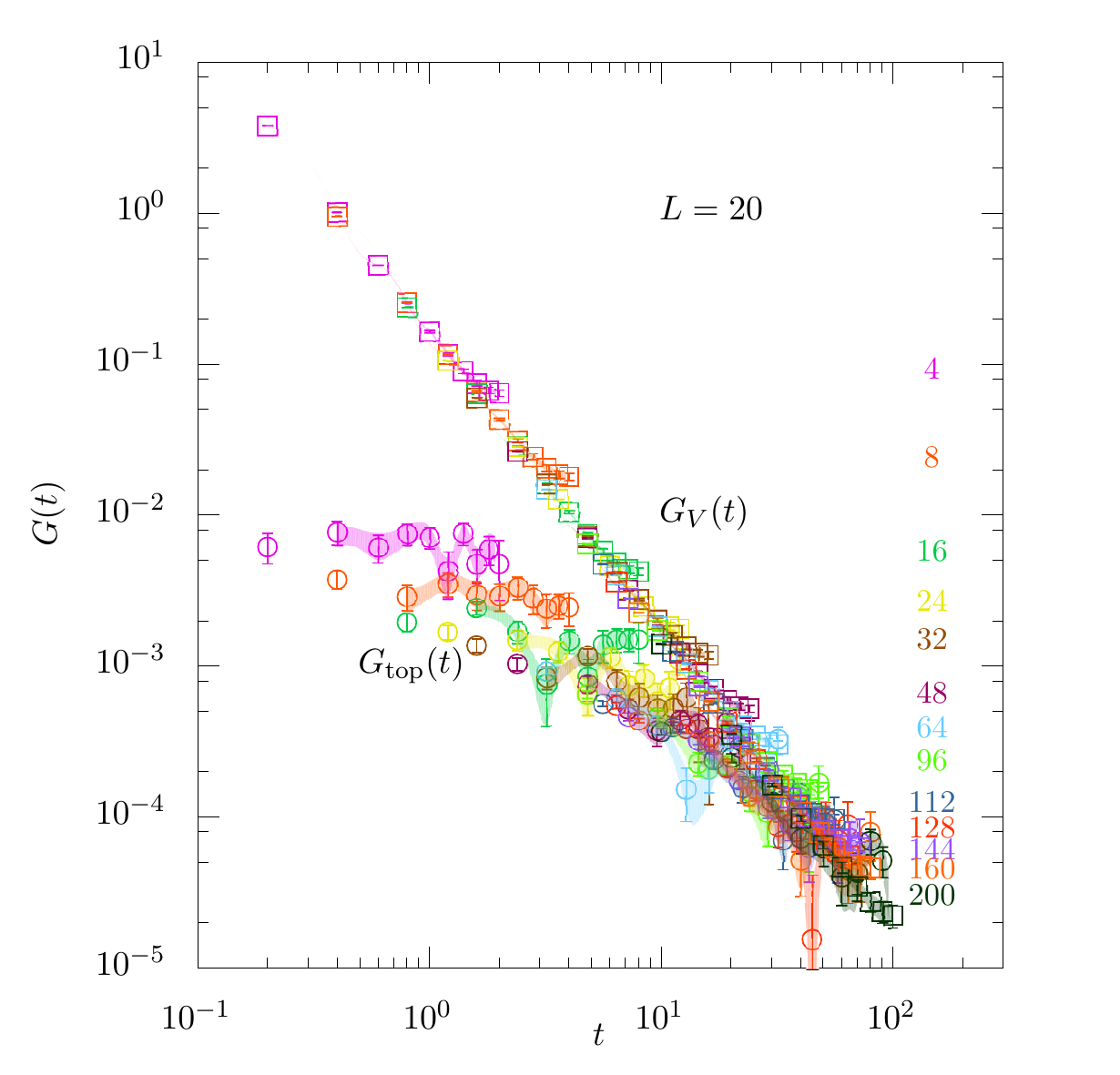}

\includegraphics[scale=0.6]{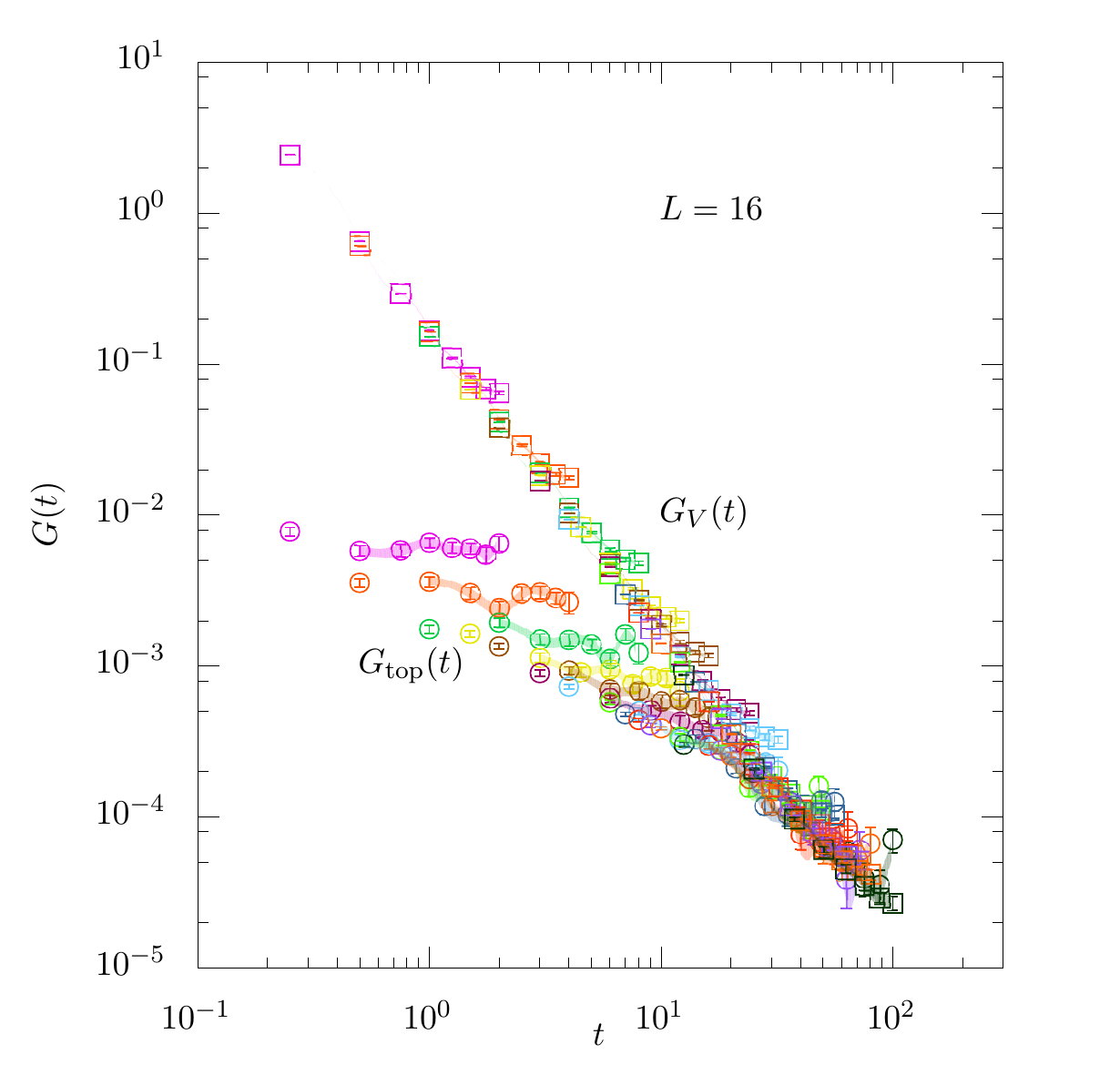}
\includegraphics[scale=0.6]{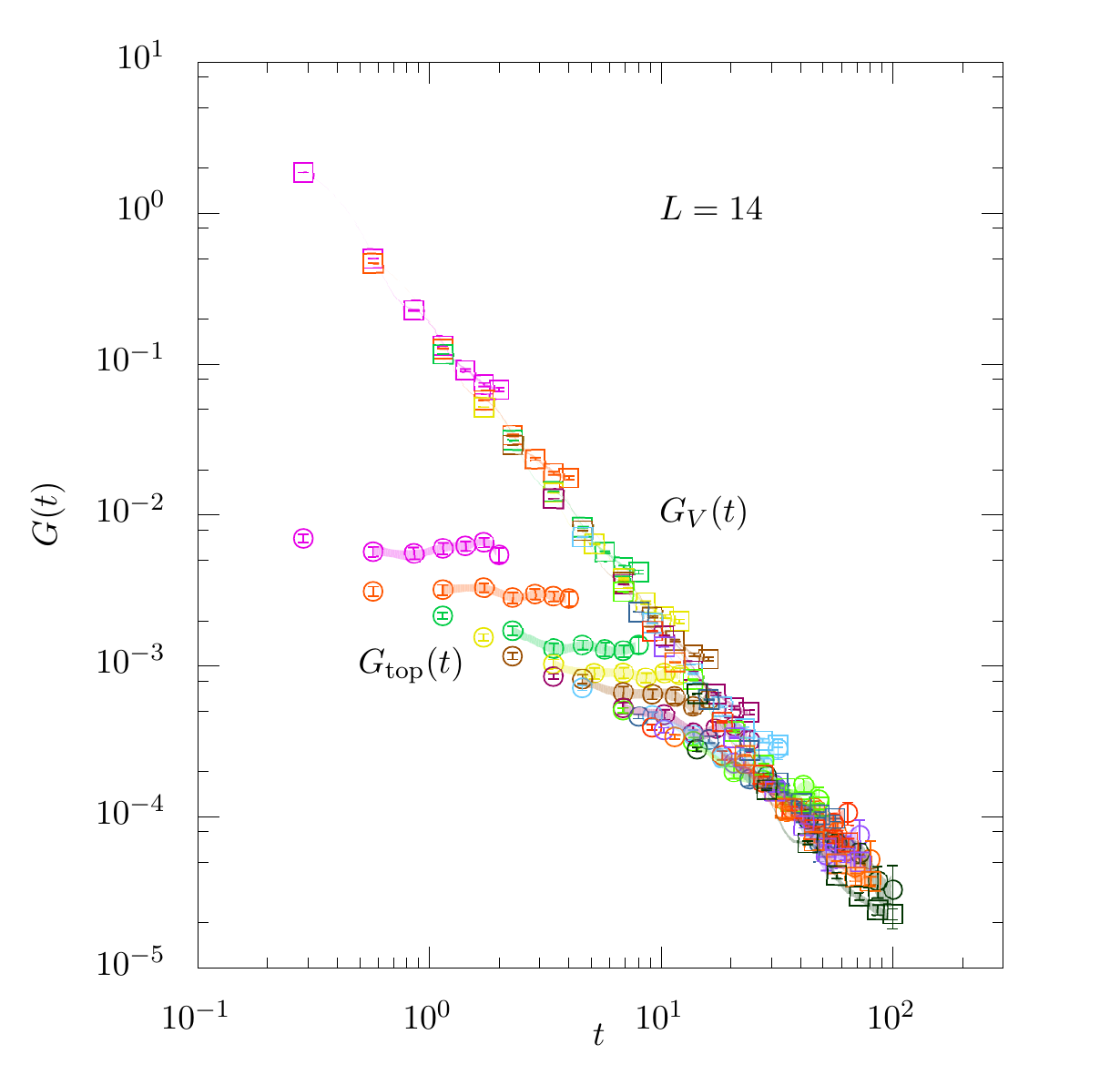}
\caption{The vector correlator $G^V(t)$ (squares) and the topological
current correlator $G_\topo(t)$ (circles) at different fixed $L$ are
shown in the four panels. The different colored symbols at each $L$
correspond to the data from different $\ell$ as specified by the color-code in the top-right panel.}
\eef{gjgt}

Arguments based on the self-duality of the two flavor massless QED$_3$
suggests that the global SU$(2)$ symmetry present in QED$_3$
Lagrangian gets enhanced into O$(4)$ symmetry at the conformal
point in the infra-red limit~\cite{Wang:2017txt}. If this is true,
the amplitude of the correlator $G_\topo(t)$ of the topological current,
\be
j_k({\bf x}) = \frac{1}{2\pi} \epsilon_{klm} \partial_l A_m({\bf x}),\qquad G_\topo(t) = \int dx dy \left\langle\sum_{k=1}^2 j_k(0,0,0) j_k(x,y,t)  \right\rangle,
\ee
has an asymptotic behavior given by~\footnote{The scaling dimension
of this operator is same as the vector bilinear~\cite{Hermele:2005dkq}.}
\be
G_\topo(t) =\frac{ C_J^t(t=\infty)}{t^2};\qquad {\rm as}\quad t\to\infty,\label{gtasymp}
\ee
and we expect 
\be
C_J^t(t=\infty) = C_J^f(t=\infty).\label{o4sym}
\ee
This is a non-trivial check since this correlator is trivial in the
pure gauge theory where there is no dependence on the separation $t$.
However, the computation using Feynman diagrams for QED$_3$ with a large number
of flavors~\cite{Giombi:2016fct} yields~\footnote{Note that the
normalization of the vector current and the topological current
differ by a factor of $2$ in~\cite{Giombi:2016fct,Chester:2016ref}
but we have normalized both currents by $C_J(0)$.}
\be
\frac{C^t_J(\infty)}{C^f_J(0)} =  \frac{3.3423}{N} - \frac{0.4634}{N^2} + \mathcal{O}\left(\frac{1}{N^3}\right).\label{giombit}
\ee
For $N=2$, the value is 1.55. This result is mildly 
different from the value 1.07 from \eqn{giombiv} implying that the large $N$ calculation does not predict
enhanced O$(4)$ symmetry for $N=2$. This is not surprising since \eqn{giombit} is strictly
valid only for large $N$ and the equality of the two amplitudes is
expected only for $N=2$.

\bef
\centering
\includegraphics[scale=0.7]{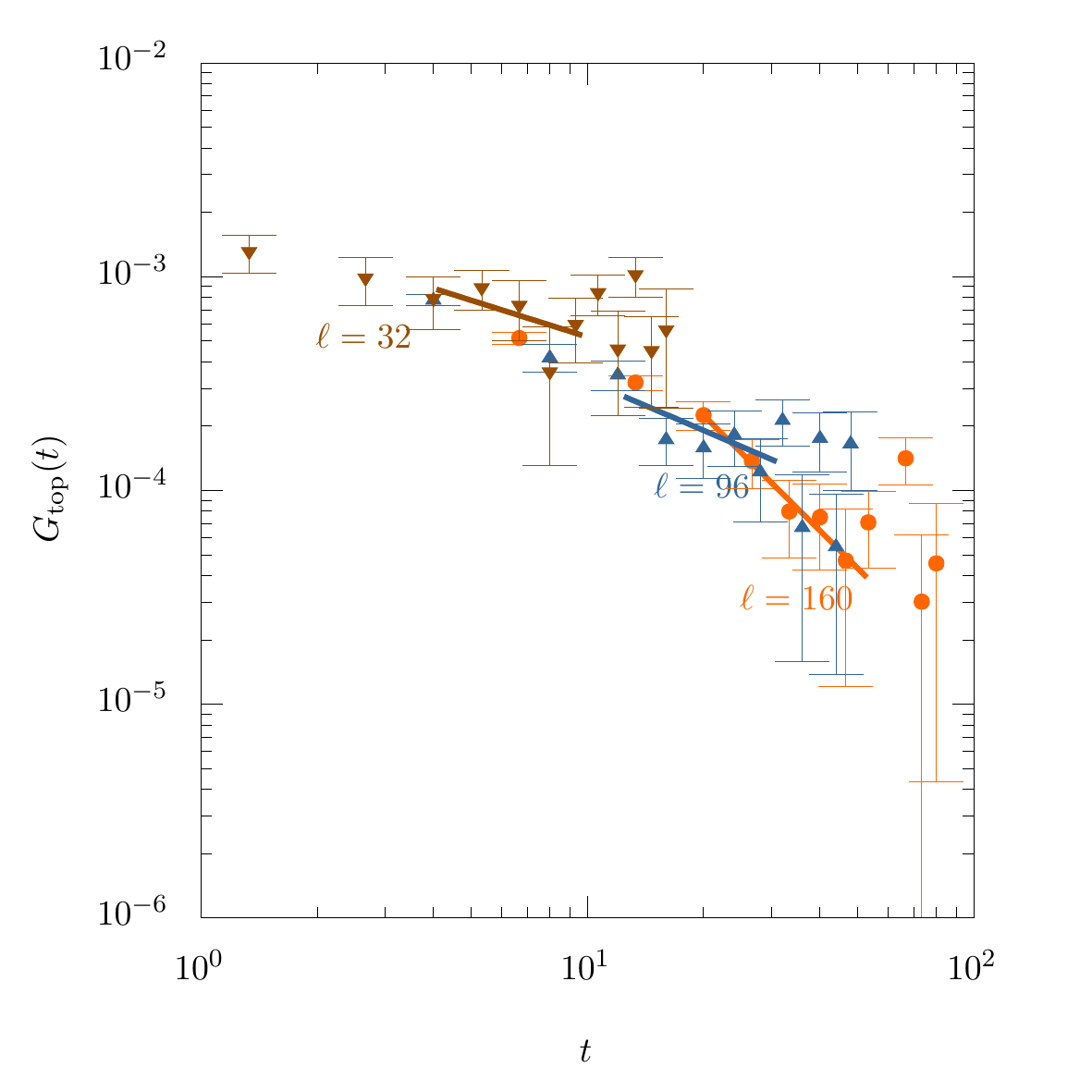}
\includegraphics[scale=0.7]{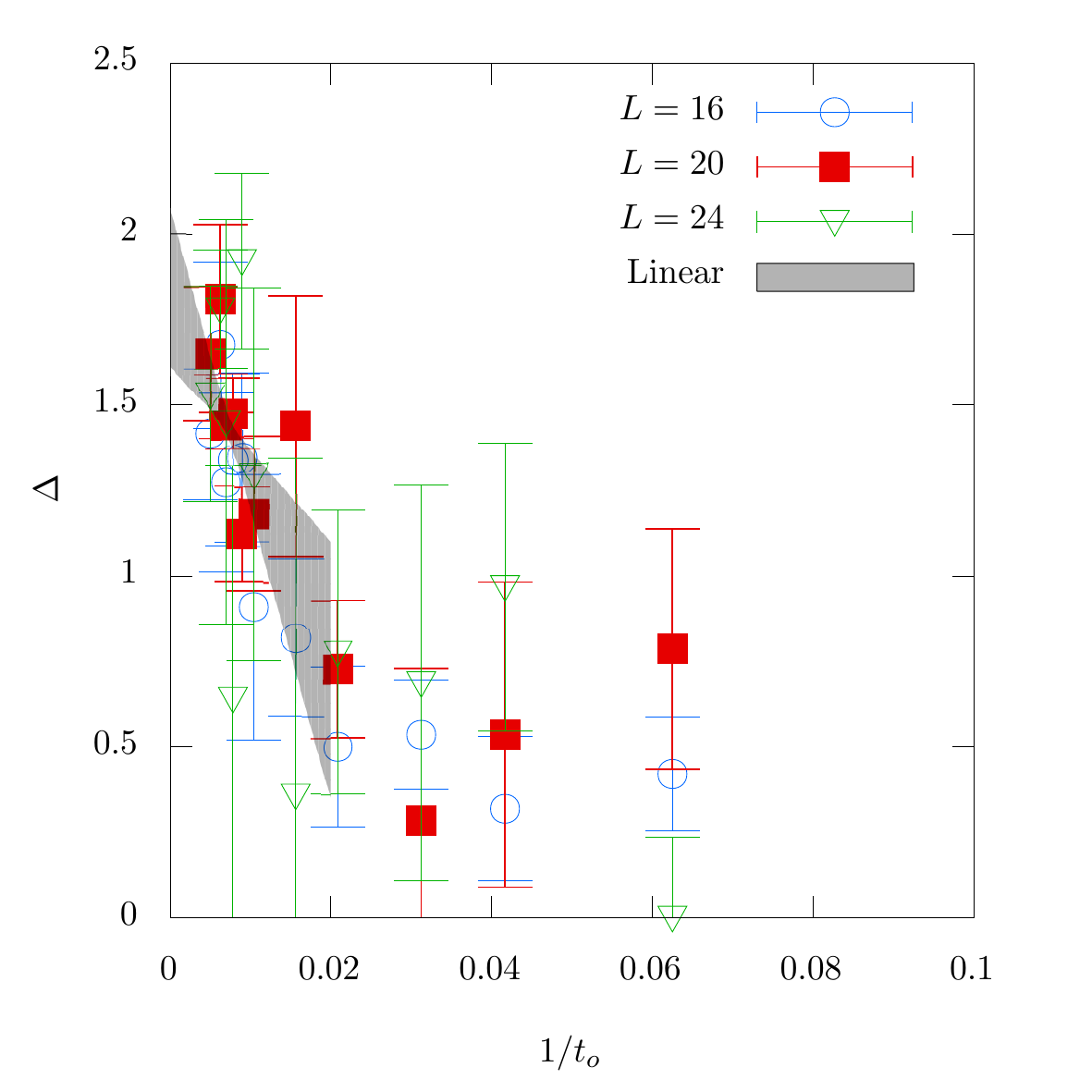}
\caption{
(Left) Power-law fits $G_{\topo}(t)\sim t^{-\Delta}$, shown as solid straight lines, using data in the range 
$t_o\le t \le 2 t_o$. The different colored symbols correspond to $G_\topo(t,\ell)$ from different $\ell$. 
We chose $t_o=\ell/8$.
(Right) The increase of the exponent $\Delta$ towards the value 2 as $t_o$ becomes larger is shown. 
The different colored symbols correspond to different lattice sizes $L$. The gray band is a linear $A+B/t_o$ fit 
to the $L=20$ data.
}
\eef{topdim}

On the lattice, we determined the topological current correlator as
\bea
j^{\rm \scriptscriptstyle lat}_k(T)&=&\frac{1}{2\pi}\sum_{X,Y=1}^L\sum_{l,m} \epsilon_{klm}\left(\theta_m(\mathbf{X}+\hat{l})-\theta_m(\mathbf{X})\right);\qquad\cr G_\topo(T)&=&\frac{1}{\ell^2}\left\langle \sum_{k=1}^2 j^{\rm \scriptscriptstyle lat}_k(T)j^{\rm \scriptscriptstyle lat}_k(0)\right\rangle
\label{lattop}
\eea
where $\theta_k(\mathbf{X})=A_k(\mathbf{X})\ell/L$ is the (unsmeared)
lattice gauge field from the lattice site $\mathbf{X}=(X,Y,T)$.
Also, we have projected $j_k$ to zero momentum at both the source
and sink time-slices in order to improve the signal, and then divided
by $1/\ell^2$ to obtain the
topological current correlator at zero spatial momentum. We do not
use $j_3$ in the analysis since its integral over the $xy$-plane
is zero for the non-compact gauge field we use. Also, the definition
in the second line of \eqn{lattop} is consistent with the definition
of the flavor-triplet vector bilinear correlator in \eqn{vecdef}.

\bef
\centering
\includegraphics[scale=0.6]{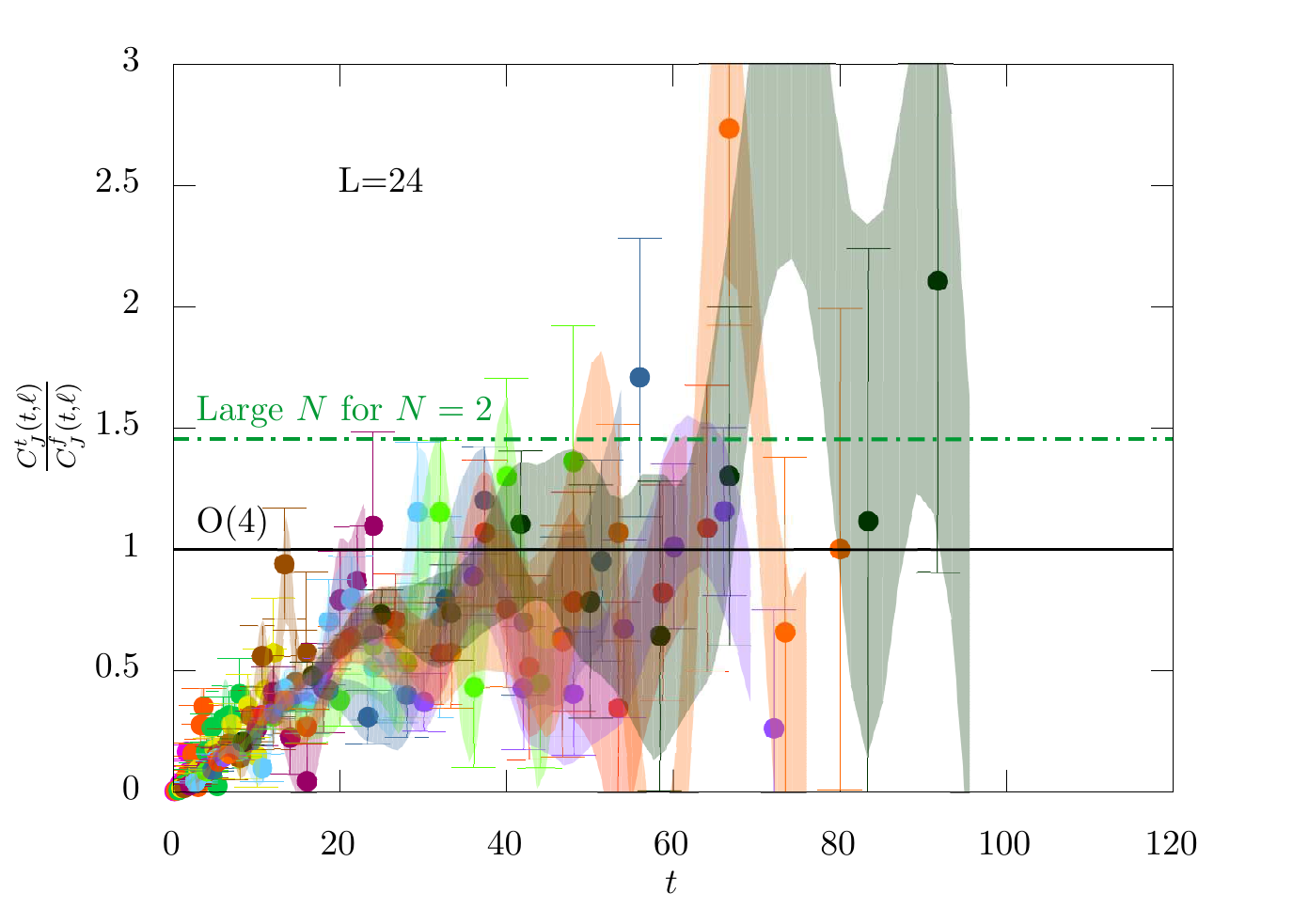}
\includegraphics[scale=0.6]{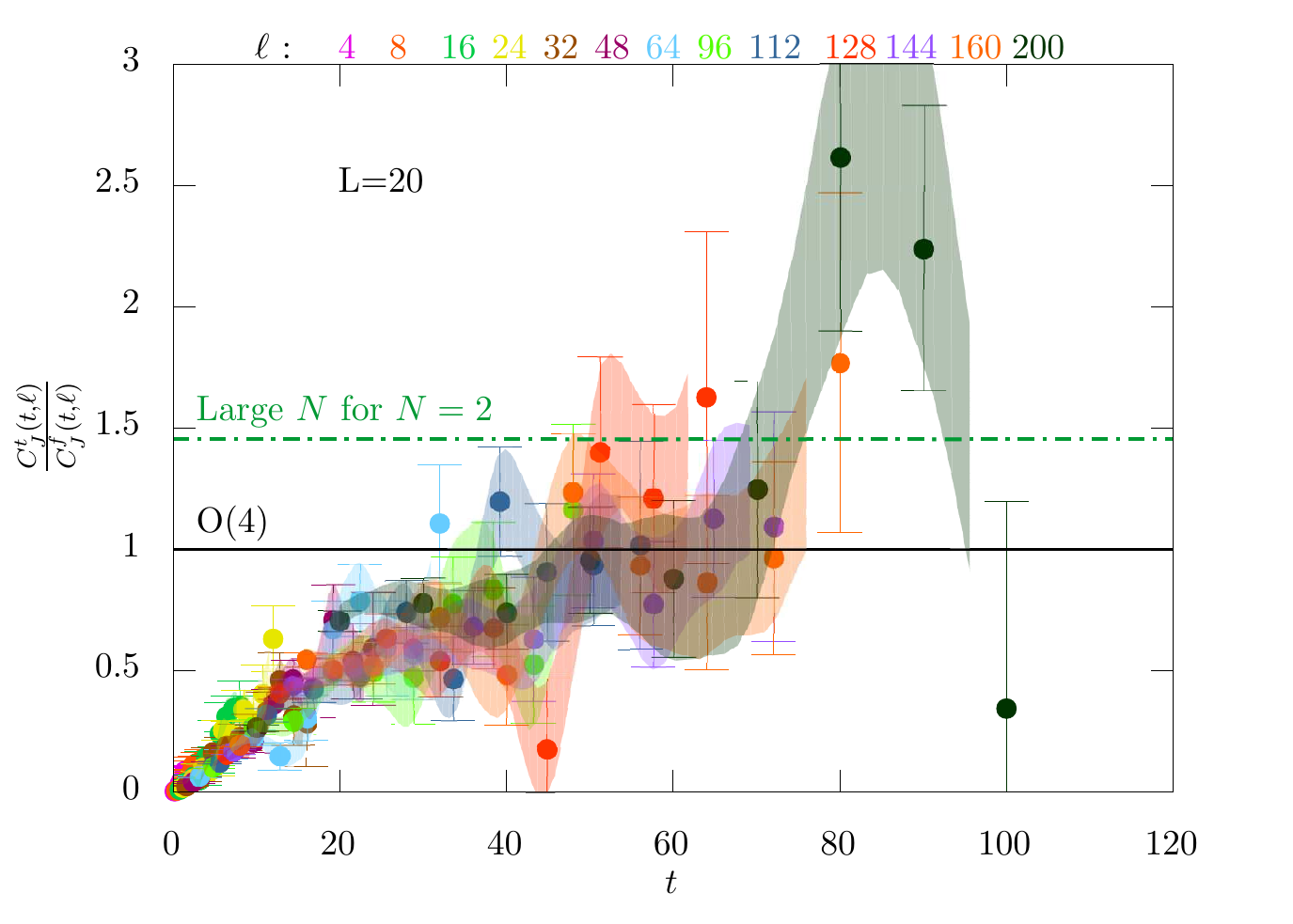}

\includegraphics[scale=0.6]{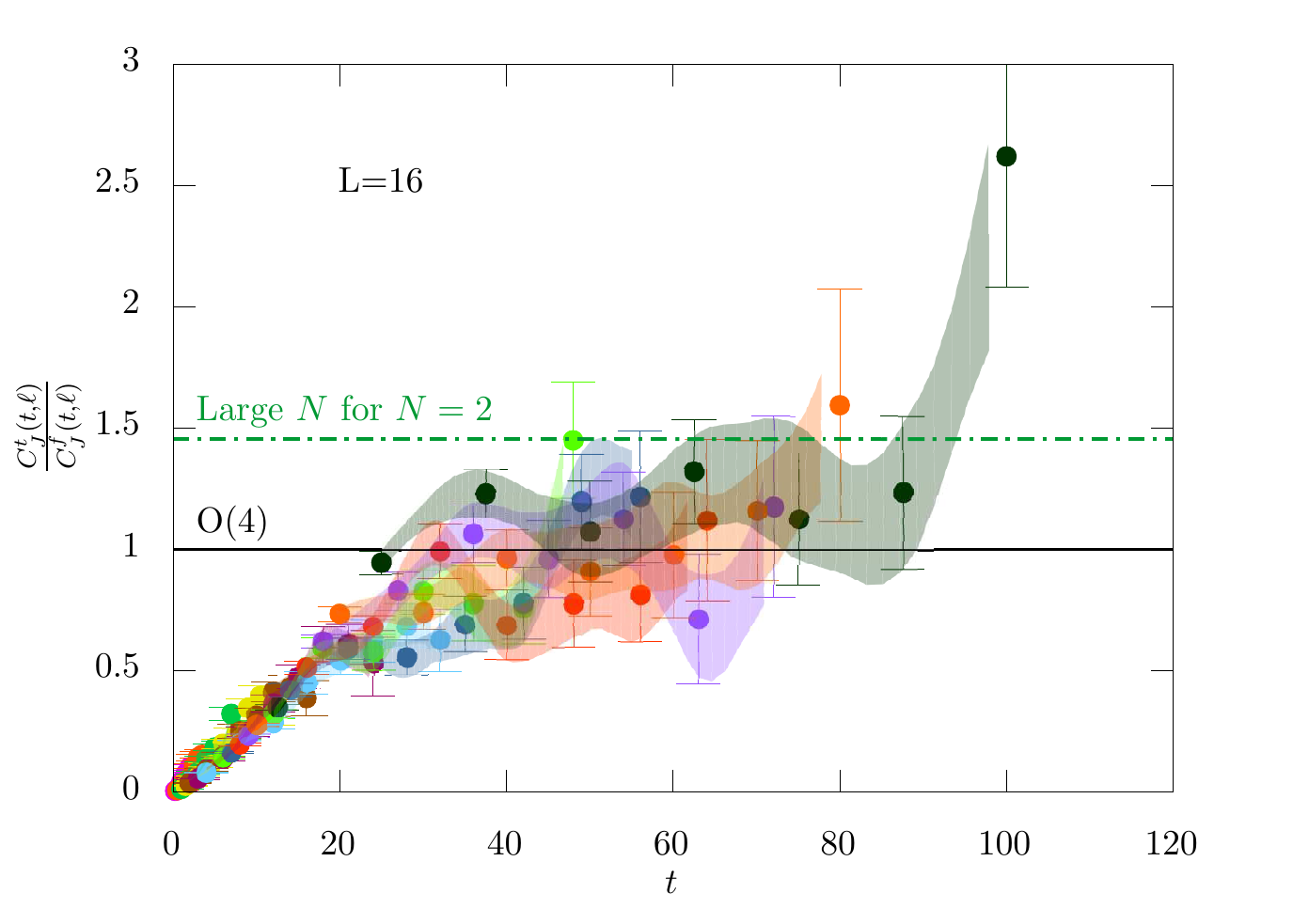}
\includegraphics[scale=0.6]{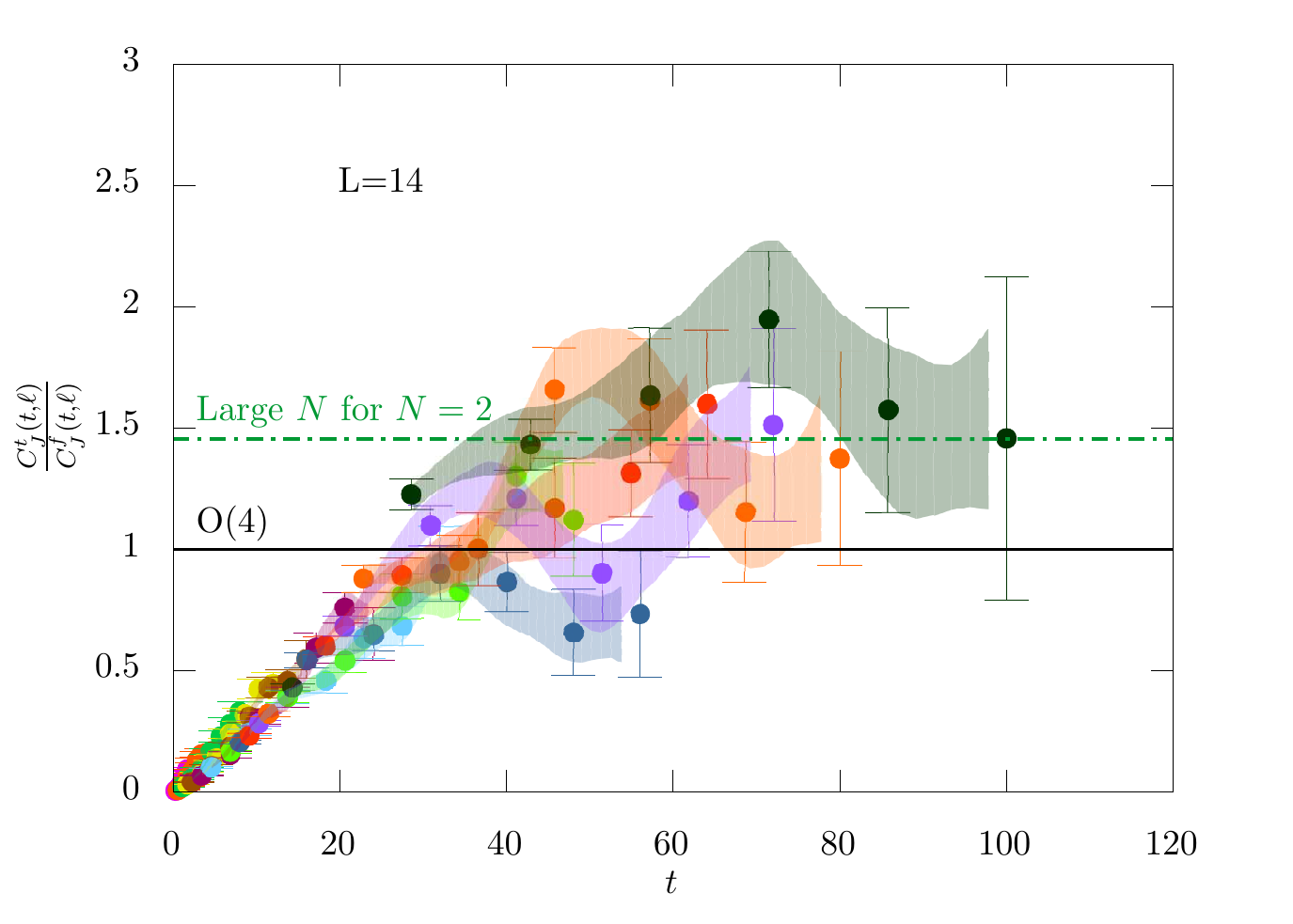}
\caption{The ratio $\frac{C^t_J(t,\ell)}{C^f_J(t,\ell)}$ are shown as
a function of $t$ at different fixed $L$ in the four panels. The
different colored symbols and bands are the data and their cubic
spline interpolations at different $\ell$. The color code for $\ell$ is shown on top
of the top-right panel. The expectation for this
ratio from the O$(4)$ symmetry is 1 as $t\to\infty$. The dot-dashed
line is the result from the large-$N$ computation extended to $N=2$.}
\eef{ratio}

The results for $G_\topo(t)$ are compared with $G_V(t)$ in \fgn{gjgt}.
We have used the data from different $\ell$ at same $L$ in order
to span a range of $t$, as explained in the last section. The
different colored symbols in each of the four panels in \fgn{gjgt}
correspond to different $\ell$.  A detailed analysis of the type
performed in the previous section does not work here due to larger
errors in the topological current correlator, which is a pure-gauge
observable, compared to the fermionic vector current correlator.
This lead to uncontrolled errors when we attempted the $L\to\infty$
and $\ell\to\infty$ extrapolations, especially at large values of
$t$ where we are interested. Therefore, we restrict ourselves to
comparisons on finite lattices at different $\ell$.

Unlike $G_V(t)$ which is a correlator of a conserved current, the
behavior of $G_\topo(t)$ is not a simple power law for all values
of $t$.  At small values of $t$, $G_\topo(t)$ is orders of magnitude
smaller than that of $G_V(t)$.  The propagator has to be monotonic
in $t$ and if it were to have a non-zero limit for every $t$ as
$\ell\to\infty$, then our data suggests that the propagator approaches
a non-zero constant at these short distances.~\footnote{We cannot
rule out the possibility that this propagator has a trivial
$\ell\to\infty$ limit for all $t$. We assume this is not the case.}
As $t$ becomes larger ($t>10$), $G_\topo$ is seen to approach $G_V$.
Our data at all values of $L$ show reasonably good evidence for a
region in $t$ where the correlators $G_V(t)$ and $G_\topo(t)$ match.
The errors in $G_\topo$ get worse as $L$ increases due to a decrease
in statistics associated with an increase in autocorrelation in the
simulation when $L$ is increased.

\bef
\centering
\includegraphics[scale=0.8]{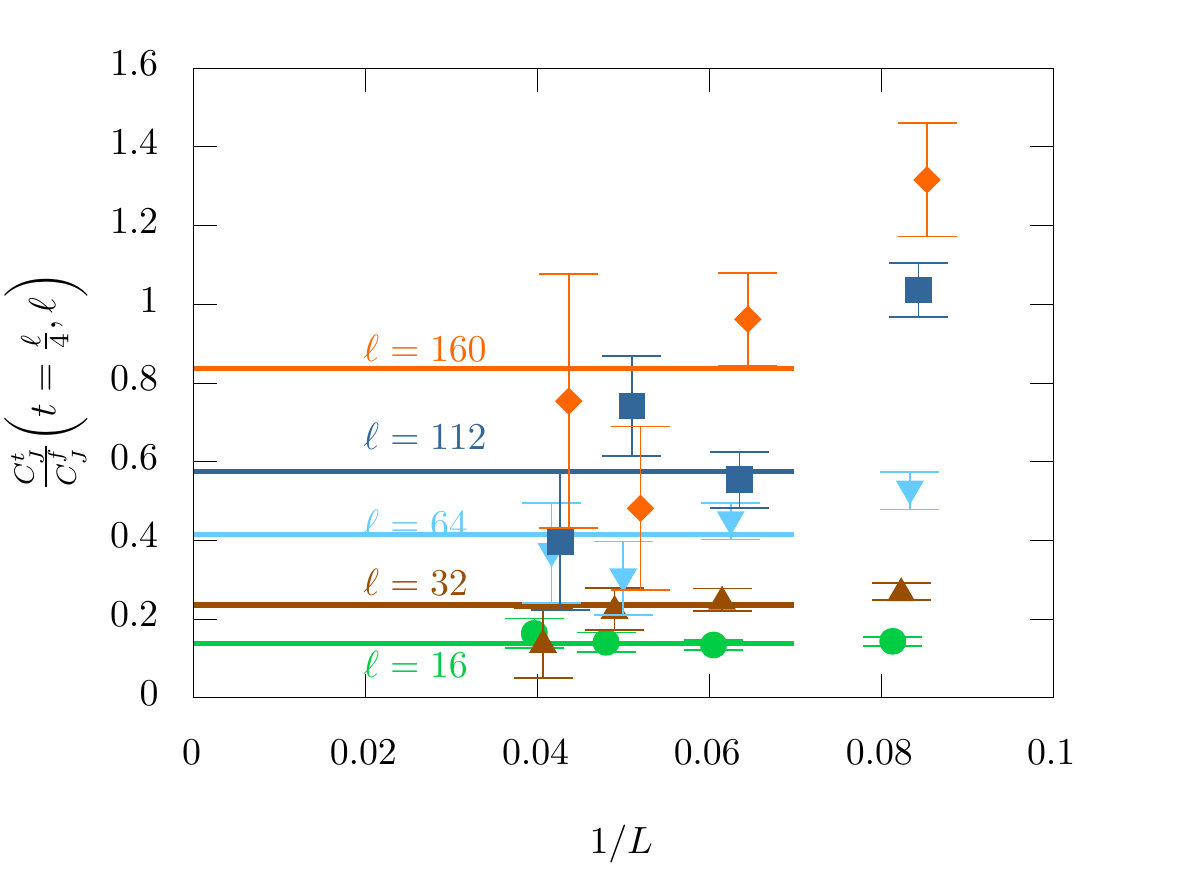}
\caption{
Lattice spacing effects in $\frac{C_J^t}{C_J^f}$ at $t=\ell/4$ data
point as determined using correlators at different $\ell$. The values at $L=16,20$ and $24$ seem
consistent within error bars.  However, at larger $\ell$ at finite
$L$, the error bars on the data also get larger to mask lattice
spacing effects.
}
\eef{toplat}

The degeneracy of the current correlators
requires that $G_\topo(t)\sim t^{-2}$ for
large $t$. To verify this, we fit a power-law $G_\topo(t,\ell)\sim
t^{-\Delta}$ to the correlators determined in finite physical volume
$\ell^3$, using data that lie in a range $t_o\le t \le 2 t_o$. We find a
reasonable power-law behavior when we choose the range corresponding
to $t_o=\ell/8$ --- a reason could be that the finite
volume effects at $t\approx\ell/2$ are avoided, and finite $L$ effects
at even smaller $t/\ell$ are also avoided. Such sample power
law fits for the correlators at $\ell=32,96$ and 160 on $L=24$
lattice are shown in the left panel of \fgn{topdim}. On the right
panel of \fgn{topdim}, we show the exponent $\Delta(t_o)$ so
determined, as a function of $1/t_o$ at three different lattice
sizes $L=16,20$ and 24. There is evidence at all three $L$ that
$\Delta$ approaches the expected value $2$ in the $t_o\to\infty$
limit.

To further explore the comparative behavior of $G_V(t)$ and
$G_\topo(t)$ at large $t$, we have plotted their ratio
\be
\frac{C_J^t(t,\ell)}{C_J^f(t,\ell)}\equiv\frac{G_\topo(t,\ell)}{G_V(t,\ell)},
\ee
in \fgn{ratio} at four different $L$ shown in the four panels. We
have shown the ratio obtained from \eqn{giombiv} and \eqn{giombit}
for comparison.  Within errors, the results at all values of $L$
are consistent with the ratio approaching unity for larger $t$ and
we see no significant difference between $L=16,20,24$ data. While
one cannot use the $L=14$ data at $t>40$ to distinguish between the
large $N$ and O$(4)$ cases, the results on finer $L=16,20,24$
lattices seem to be more consistent at the level of 1-$\sigma$ with
the O$(4)$ expectation. For $t>60$, the data becomes very noisy.
We illustrate the lattice spacing effects further in \fgn{toplat}
--- we show $\frac{C^t_J(t,\ell)}{C^f_J(t,\ell)}$ as determined at
$t=\ell/4$\ ~\footnote{In this way, interpolation can be avoided
as the data point at $t=\ell/4$ is always present on $L=12,16,20$
and $24$.} from the correlators determined in boxes of finite
physical extents $\ell$, as a function of $1/L$. The $L=16,20$ and
$24$ data are always consistent with each other as seen by the
horizontal straight lines in the figure.  Any increase in finite
lattice spacing effect as $\ell$ is increased at finite $L$ is
overcome by a corresponding increase in the noise in the topological
current correlator.  Therefore, at the level of statistical uncertainties
in \fgn{ratio}, the lattice spacing effects seem to be unimportant.

\section{Quenched ($N=0$) QED$_3$}
\bef
\centering
\includegraphics[scale=0.9]{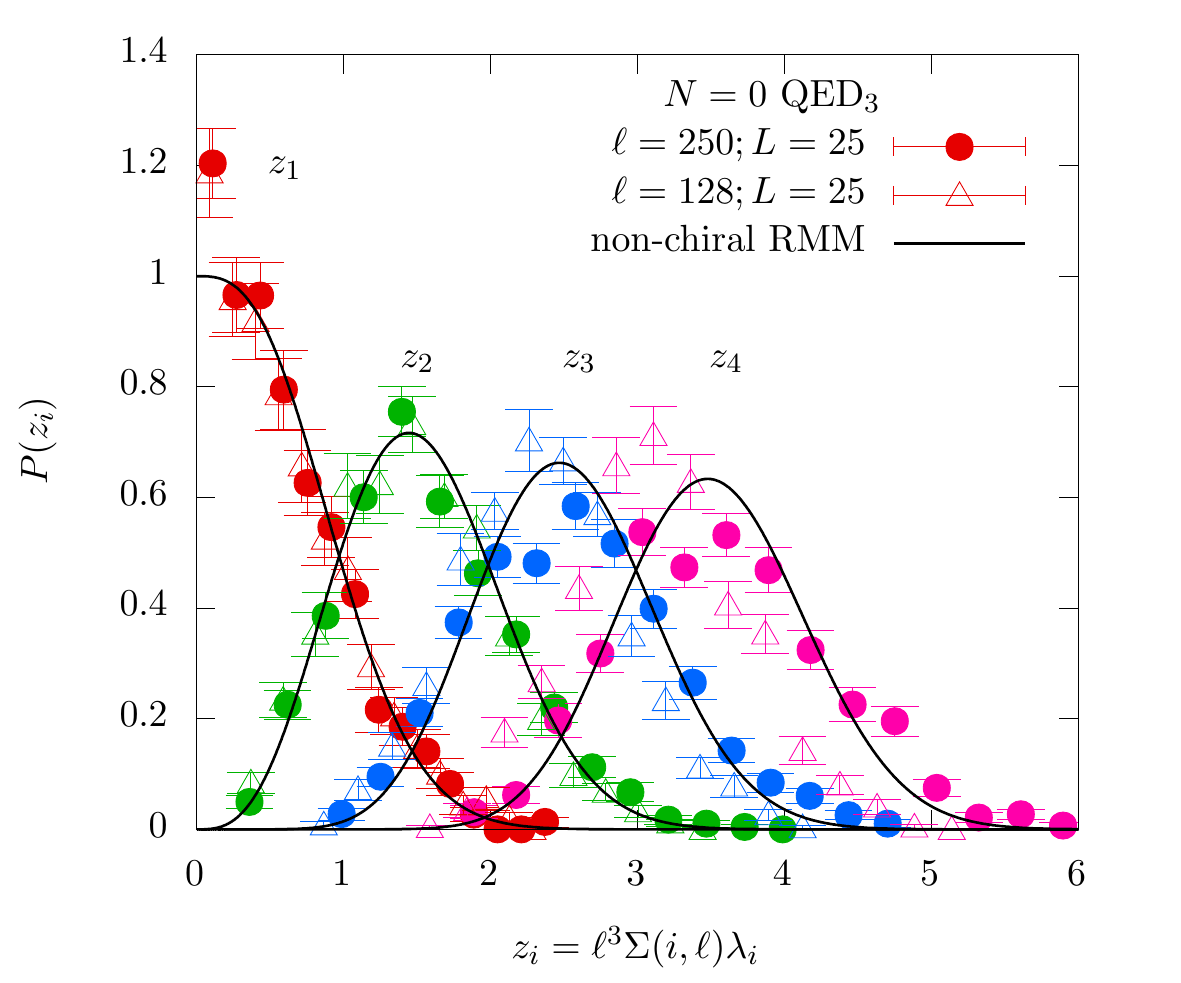}
\caption{Distribution of the low-lying eigenvalues, $\lambda_i$ for $i=1,2,3,4$, of the overlap
Dirac operator (data points) scaled by $\ell^3$. The distributions
are also multiplied by constants $\Sigma(i,\ell,L)$ such that their
means match that of the corresponding distributions of the low-lying
eigenvalues from the non-chiral random matrix model with $N=0$
(solid black curves).  Further, in the $\ell\to\infty$, all the
$\Sigma(i,\ell,L)$ have to approach the same value independent of
$i$.}
\eef{dist}

Unlike QED${}_3$ with dynamical fermions, we expect the quenched
theory where the fermions are used as a probe to have a non-conformal infra-red behavior 
with a scale set by the gauge coupling. We will assume a non-compact action for the
gauge field and therefore monopoles will be suppressed. As in our
previous paper~\cite{Karthik:2016ppr}, we study the low lying
microscopic eigenvalues, $i\lambda_j$, of the anti-Hermitian massless
overlap Dirac operator.  The presence of a bilinear condensate
implies a non-zero density at zero eigenvalue and level repulsion
implies that the level spacing of eigenvalues near zero will be
inversely proportional to $\ell^3$.  The individual distributions
of the low-lying eigenvalues (ordered by their absolute values)
will be governed by an appropriate non-chiral random matrix model
(RMM)~\cite{Verbaarschot:1994ip,Szabo:2000qq}, which in our case
will be a Hermitian random matrix model.

We simulated the quenched $N=0$ QED$_3$ by Monte Carlo sampling of
the Fourier modes of the gauge field.  We used lattices with
$L=15,17,19,21$ and 25 in order to take the continuum limit at
different $\ell$.  On the random matrix side, the distributions of
the low lying eigenvalues $z_j$ in the RMM model can be obtained
by using the sinc-kernel and the associated Fredholm
determinants~\cite{Mehta:2004,Nishigaki:2016nka}.  We numerically
evaluated the eigenvalues of the kernel required for the computation
of the determinants and traces of the resolvents, and we were able
to determine the distributions of the five lowest eigenvalues $z_j$
in the RMM needed for our comparison to a very good accuracy.

\bef
\centering
\includegraphics[scale=0.9]{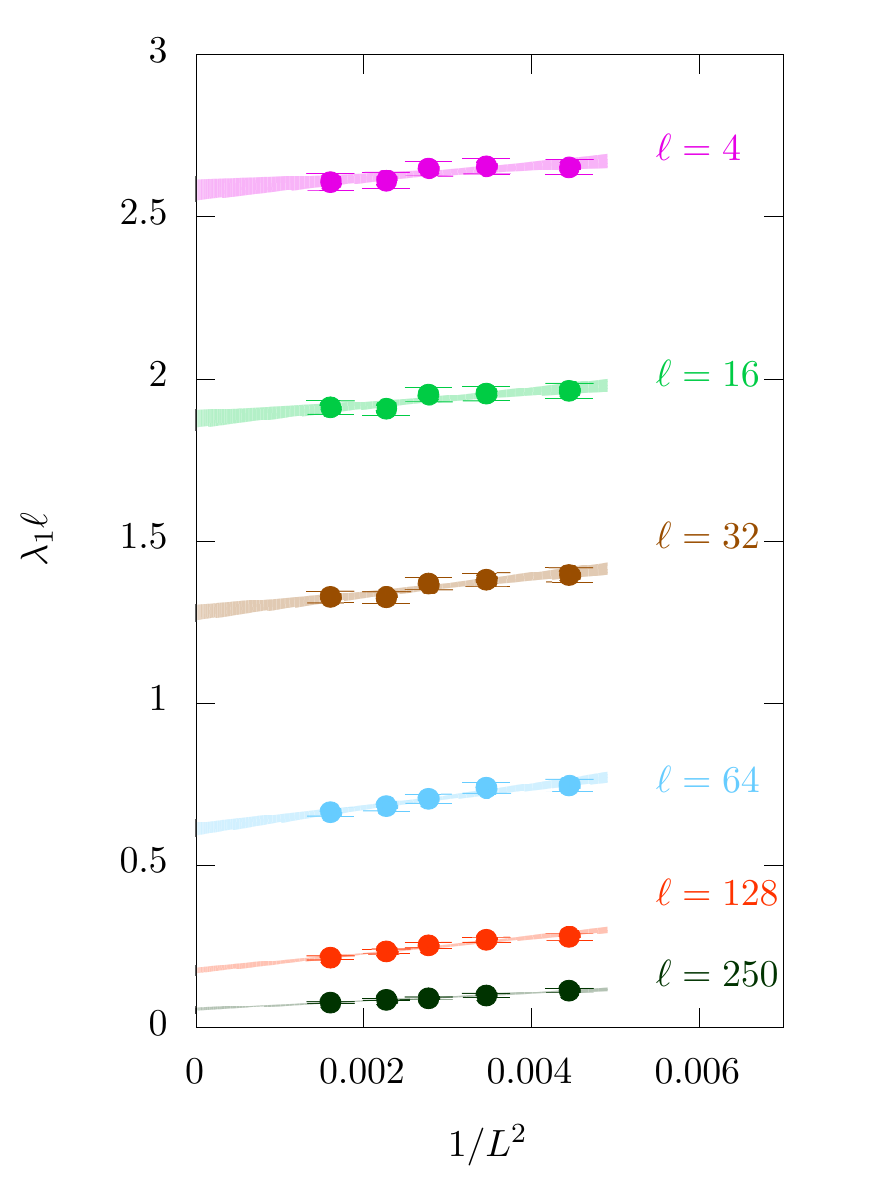}
\includegraphics[scale=0.9]{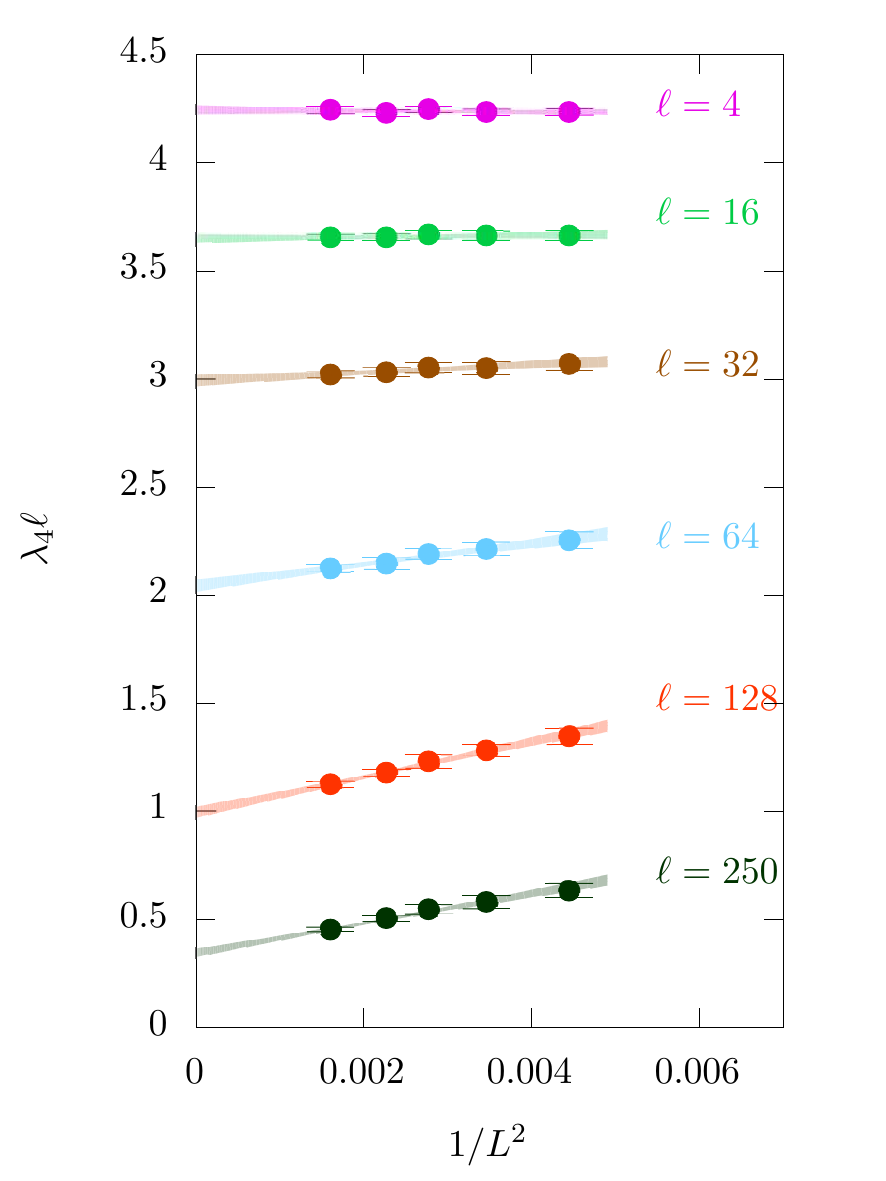}
\caption{Continuum extrapolation of first (left panel) and 
the fourth (right panel) smallest eigenvalues of the overlap Dirac 
operator. The points of different colors are the data at different fixed $\ell$. 
The bands are the $1/L^2$ extrapolation of the data.}
\eef{eigcont}

The bilinear condensate, if present, can be obtained by matching
the distribution of the low-lying microscopic eigenvalues in the
pure gauge theory to that from the RMM model. In \fgn{dist}, we
make such a comparison by scaling $\lambda_i(\ell,L)$ by a
constant $\Sigma(i,\ell,L)\ell^3$ such that the means of the two
distributions match \ie,
\beq
\Sigma(i,\ell,L) = \frac{\langle z_i\rangle}{\left\langle \lambda_i(\ell,L)\right\rangle\ell^3}.
\eeq{sigdef}
A good agreement is seen between the distributions till the
4th eigenvalue just by this simple scaling. The agreement gets
better as $\ell$ is increased as expected when a condensate is
present. In the $\ell\to\infty$ limit, taken after the $L\to\infty$
continuum limit, the values of $\Sigma$ obtained from the different
microscopic eigenvalues have to be the same, and it is the value
of the condensate. We now proceed to show this to be the case and
obtain the value of $\Sigma$.

\bef
\centering
\includegraphics[scale=0.9]{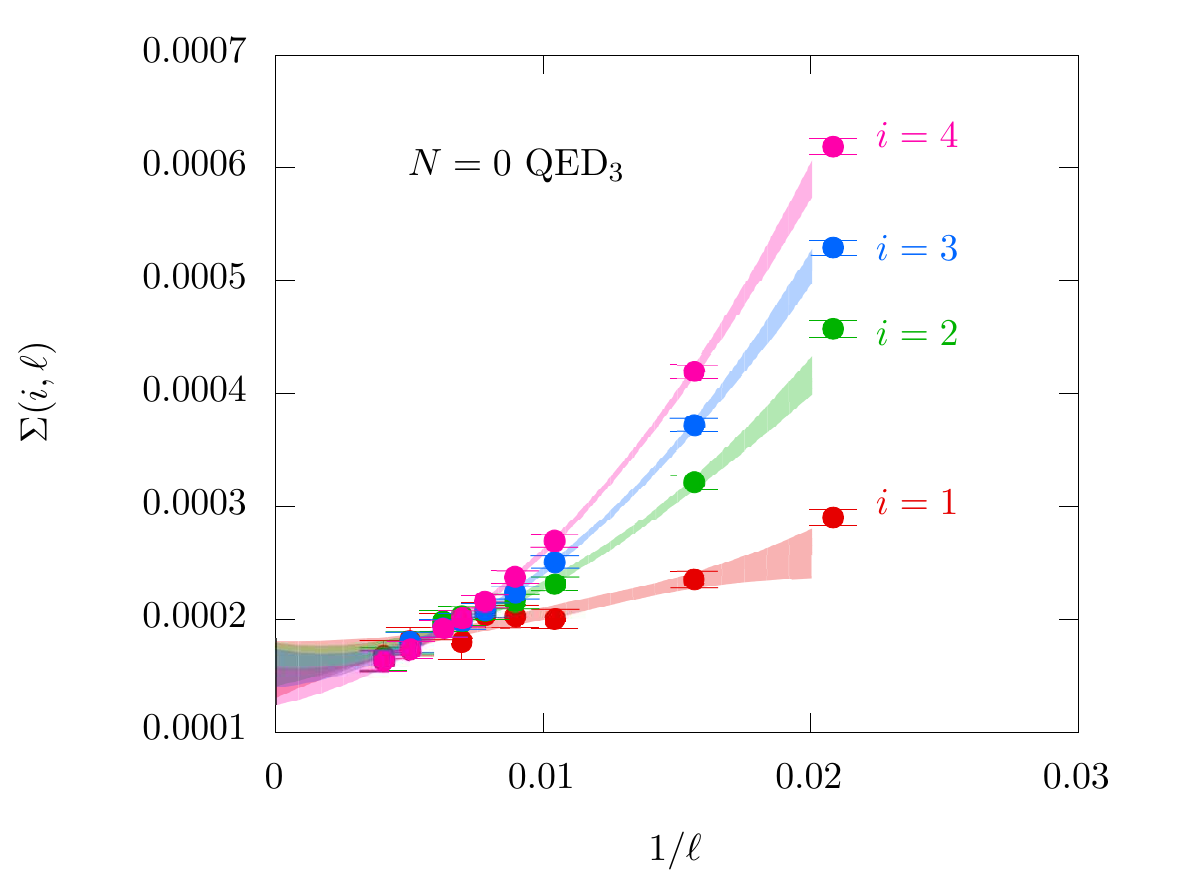}
\caption{The infinite $\ell$ limit of the condensate
$\Sigma_i(i,\ell)=\frac{\langle
z_i\rangle}{(\left\langle\lambda_i(\ell)\right\rangle\ell^3}$, where
$\langle\lambda_i(\ell)\rangle$ is the $i$-th eigenvalue of the
overlap Dirac operator after taking the $L\to\infty$ continuum
limit, and the $z_i$'s are the eigenvalues of the non-chiral RMM. The
1-$\sigma$ bands for an $A+B/\ell+C/\ell^2$ extrapolation of the finite
volume data are shown. Agreement between different $\Sigma(i,\ell=\infty)$
are seen, thereby ensuring the self-consistency of the random matrix
analysis. We estimate the condensate to be $\Sigma=1.5(1)\times10^{-4}$.
}
\eef{sigma}

We extrapolate $\langle\lambda_i(\ell,L)\rangle\ell$ to the continuum
by using a fit of the form
$\left\langle\lambda_i(\ell,L)\right\rangle=\left\langle\lambda_i(\ell)\right\rangle\ell+k/L^2$ 
at each fixed finite box size $\ell$. We show this extrapolation
at different $\ell$ for the first and the fourth smallest eigenvalues
on the left and right panels of \fgn{eigcont} respectively. Using
these continuum extrapolated values of $\langle\lambda_i(\ell)\rangle\ell$, we
determined the values of $\Sigma(i,\ell)$ from \eqn{sigdef}. The
dependence of $\Sigma(i,\ell)$ on $\ell$ for the first four eigenvalues
are shown in \fgn{sigma}.  A strong dependence on $\ell$ is seen.
However, one can easily see that they approach a non-zero limit as
$\ell\to\infty$. We extract this limit from different $i$th eigenvalues from a
$\Sigma(i,\ell)=\Sigma(i)+a_1/\ell+a_2/\ell^2$ extrapolation using
the data at $\ell\ge64$. The value of the condensate
for $i=$1,2,3 and 4 are $1.5(2)\times10^{-4}$, $1.6(2)\times10^{-4}$,
$1.6(2)\times10^{-4}$ and $1.4(2)\times10^{-4}$ respectively. They are all 
consistent with each other thereby assuring the consistency of the method. 
Taking their average, we estimate the value of the condensate in quenched QED$_3$
to be $1.5(1)\times10^{-4}$. For comparison, the value of the condensate per color degree of 
freedom in the 't Hooft limit is $4.2(4)\times 10^{-3}$~\cite{Karthik:2016bmf}.

\section{Conclusions}

A further study of the correlator of the flavor-triplet vector
bilinear in QED${}_3$ with two flavors of two component massless
fermions suggests an enhanced O$(4)$ symmetry in the infra-red limit
as predicted by a strong duality~\cite{Wang:2017txt}.  The amplitude
of the correlator of the flavor-triplet vector bilinear $C_J^f$ and
the amplitude of the correlator of the topological current $C_J^t$
are the same in the large distance limit in our numerical calculation.
There is an intermediate region in the separation where the amplitude
$C_J^f$ itself is lower than its ultraviolet value and it is likely
that this trend remains as one approaches the infrared limit. A
further check on whether the enhanced O$(4)$ symmetry in $N=2$
QED$_3$ also implies its duality to the easy plane $NCCP^1$ model
proposed in~\cite{Wang:2017txt} will involve a computation of the
scaling dimensions of certain four Fermi operators. We plan to
address this along with the behavior of other higher dimensional
composite operators in the future.

We show clear evidence for a bilinear condensate in the quenched
theory -- pure gauge theory with massless fermions as a probe.  Our
results also show that the quenched theory has a finite condensate
in the infinite volume limit.  This is contrary to what happens in
even dimensions~\cite{Damgaard:2001xr,Damgaard:2005wz} where a
diverging condensate is usually associated with the presence of an
axial anomaly in the theory.  We have studied the pure gauge theory
where contributions from monopoles have been suppressed.  It would
be interesting to see if the condensate would diverge in a theory
with a compact gauge action.  Compact gauge action poses a technical
problem since one can have anomalously small eigenvalues of a massive
Wilson-Dirac operator that is used as a kernel for the massless
overlap Dirac operator.  Preliminary investigations suggest that
such eigenvalues are suppressed in the continuum limit at a fixed
physical volume. Therefore, it should be possible to study the
quenched theory with a compact gauge action if one improves the
gauge action and the fermion operator used as the probe.  A diverging
condensate will suggest that monopoles play a physical role in the
theory. This will also make it interesting to study QED${}_3$ with
dynamical fermions and a compact gauge action.

\acknowledgments
We would like to thank Shai Chester, Igor Klebanov, Max Metlitski,
Silviu Pufu, T. Senthil and Cenke Xu for useful discussions during
a workshop at the Princeton Center for Theoretical Science.  All
computations in this paper were made on the JLAB computing clusters
under a class C project.  The authors acknowledge partial support
by the NSF under grant number PHY-1515446.

\bibliography{biblio}
\end{document}